\documentclass[letterpaper]{emulateapj}
\usepackage{epstopdf}
\usepackage{placeins}
\usepackage{natbib}
\usepackage{verbatim}

\newcommand{\civ}{\ion{C}{4}\ }
\newcommand{\ha}{H$\alpha\ $}
\newcommand{\hbeta}{H$\beta\ $}
\newcommand{\oii}{\ion{O}{2}\ }
\newcommand{\oiii}{\ion{O}{3}\ }
\newcommand{\mgii}{\ion{Mg}{2}\ }
\newcommand{\feii}{\ion{Fe}{2}\ }
\newcommand{\hh}{$h^{-1}_{70}$}

\begin{document}

\title{Observations of \mgii absorption near \lowercase{$z\sim 1$} galaxies \\ selected from the DEEP2 redshift survey\altaffilmark{1}}
\author{Elizabeth Lovegrove\altaffilmark{2,3}, Robert A. Simcoe\altaffilmark{2,4}}
\altaffiltext{1}{This paper includes data gathered with the 6.5 meter Magellan Telescopes located at Las Campanas Observatory, Chile.}
\altaffiltext{2}{MIT-Kavli Center for Astrophysics and Space Research}
\altaffiltext{3}{University of California, Santa Cruz}
\altaffiltext{4}{Sloan Foundation Research Fellow}

\begin{abstract}
We study the frequency of \mgii absorption in the outer haloes of
galaxies at $z=0.6-1.4$ (with median $z=0.87$), using new spectra
obtained of ten background quasars with galaxy impact parameters of
$b<100$ kpc.  The quasar sightlines were selected from the SDSS DR6
QSO catalog based on proximity to galaxies in the DEEP2 redshift
survey.  In addition to the 10 small impact systems, we examine 40
additional galaxies at $100<b<500$ kpc serendipitously located in the
same fields.  We detect \mgii absorbers with equivalent width $W_r$ =
0.15 \AA\ - 1.0\AA, though not all absorbers correlate with DEEP
galaxies.  We find five unique absorbers within $\Delta v = 500$ km/s
and $b<100$ kpc of a DEEP galaxy; this small sample contains both
early and late type galaxies and has no obvious trends with star
formation rate.  No \mgii is detected more than 100 kpc from galaxies;
inside this radius the covering fraction scales with impact parameter
and galaxy luminosity in very similar fashion to samples studied at
lower redshift.  In all but one case, when \mgii is detected without a
spectroscopically confirmed galaxy, there exists a plausible
photometric candidate which was excluded because of slit collision or
apparent magnitude.  We do not detect any strong absorbers with
$W_r>1.0$\AA, consistent with other samples of galaxy-selected \mgii
systems.  We speculate that \mgii systems with $0.3<W_r<1.0$ trace old
relic material from galactic outflows and/or the halo assembly
process, and that in contrast, systems with large $W_r$ are more
likely to reflect the more recent star forming history of their
associated galaxies.
\end{abstract}
\section{Introduction}

For roughly two decades, a connection has been observed between $\sim
L^*$ galaxies and the \mgii absorption systems seen in QSO spectra
\citep{bergeron_boisse, steidel,steidelandsargent}.  \mgii systems
trace gas in the outer haloes of these objects out to radii of $\sim 100$
kpc.  However, a standardized physical picture describing the origin
of this metal-rich material remains elusive.  At least two different
pathways have been suggested for distributing \mgii throughout galaxy
haloes.

In one scenario, the QSO absorbers represent debris from large-scale
winds that are seen in galaxy spectra and known to carry \mgii
\citep{ubiquitous2009, bond}.  These winds may help quench star
formation in nascent galaxies, providing necessary feedback in galaxy
formation models.  For example, \citet{nestor2010strong} associated
ultrastrong absorbers ($W_r \ge$ 3\AA) with starbursting galaxies and
suggested an outflow wind scenario based on velocity dispersions in
the \mgii systems.  \citet{prochtersdss2006} favored an outflow model
of \mgii systems based on evolutionary trends with redshift in the
number density of (strong) absorbers.  Some theories associate these
outflows with AGN activity rather than supernova feedback, but
differentiation between the two remains an open question.

Several authors have argued indirectly in favor of this interpretation
as well.  \citet{bouche2006anticorrelation} detected evidence that
stronger \mgii absorbers reside in lower mass galaxy haloes, and
suggested that this might result from the more vigorous star formation
occurring in the smaller, late-type galaxies.  Using various
statistical combinations of SDSS \mgii absorbers,
\citet{zibetti2007stacking} and \citet{menardo2} reported evidence of
star formation surrounding the strongest \mgii systems, in the form of
enhanced [OII] emission or bluer galaxy colors.

However, a second interpretation posits that \mgii systems are a
generic feature of galaxy haloes, with an extent and dynamical
structure that reflects gravitational processes.  Indeed the velocity
profiles of some \mgii systems appear to form an extension of galactic
rotation curves \citep{kacprzak2010sim, mgii_rotation}, albeit with
increased velocity dispersion.  \citet{chenandtinker2008} and
\citet{chen2010lowz} develop a halo-motivated model using a large set
of QSO-galaxy pairs.  These papers and \citet{gauthier} interpret the
anti-correlation between galaxy halo mass and \mgii strength in the
context of the ``cold mode'' vs. ``hot mode'' accretion models
recently advocated by \citet{keres, dekel}.  In this model, \mgii absorption
arises in cold structures that are ionized away when galaxy masses -
and hence virial temperatures - exceed a threshold value.  In a
carefully selected sample, \citet{chenandtinker2008} find absorbers
associated with both early- and late-type galaxies in normal field
proportions, and use this to argue against an interpretation where
\mgii arises exclusively in starburst galaxies.

Galaxy-absorber studies to date have primarily focused on low redshifts
$z\lesssim0.5$ where galaxy observations are most accessible
\citep{chen2010lowz}, or on targeted galaxy observations toward
sightlines pre-selected for \mgii absorption \citep{kacprzak2010sim,
  nestor2005, mgii_rotation}.  A uniform interpretation of the results
is also complicated by the different samples' varying sensitivities,
completeness estimates, and selection techniques.

In this paper, we construct a modest sample for studying \mgii-galaxy
associations at higher redshift ($0.6<z<1.4$).  To do this, we
leverage the public catalog of the DEEP2 redshift survey
\citep{deep2_1,deep2_2} and use the MagE spectrometer on Magellan to
obtain spectra of background QSOs that lie randomly near DEEP2
galaxies on the sky.
 
Since this galaxy sample is chosen without regard to morphology, it is
not biased toward any spectral type or the presence/absence of AGN;
thus, it should be appropriate for investigating the relationship
between \mgii incidence, star formation, and spectral/morphological
type.  The sample also spans a wide range of impact parameters,
allowing for an estimate of how the \mgii covering fraction scales
with galactic radius.  The DEEP sample does pre-select galaxies based
on color and magnitude and is known to preferentially select star
forming galaxies at high redshift \citep{gerke}.  However, it provides a large,
uniformly selected parent sample against which we can compare a
subsample drawn randomly based only on impact parameter.

In Section 2, we describe our sample selection and observations.
Section 3 details the galaxy-\mgii pairs detected as well as
non-detections.  Section 4 describes our results in the context of the
existing literature.  We use a WMAP7+BAO+$H_0$ cosmology of $\Omega_m
= 0.272, \Omega_{\Lambda} = 0.728, H_0 = 70.4$ km s$^{-1}$ Mpc$^{-1}$
throughout.  All magnitudes are on the AB system, and impact
parameters are listed in proper $h_{71}^{-1}$ kpc.

\section{Observations}

Our objective is to assemble a sample of galaxy-absorber pairs at
moderate redshift where the galaxy sample is {\em not} selected based
on pre-existing knowledge of the presence of \mgii absorption.  This
method has been employed successfully at lower redshifts but becomes
somewhat less practical at higher $z$ because of the expense and
incompleteness of the galaxy observations.  To work around this
limitation, we started with the well-defined, public galaxy redshift
catalog of the DEEP2 survey, which contains galaxies at $0.6<z<1.4$.
We then selected for observation all quasars in the SDSS DR6 catalog
whose sightlines passed less than $100$ kpc from a DEEP galaxy.  All
quasars meeting this criterion were observed.

\begin{figure}
\plotone{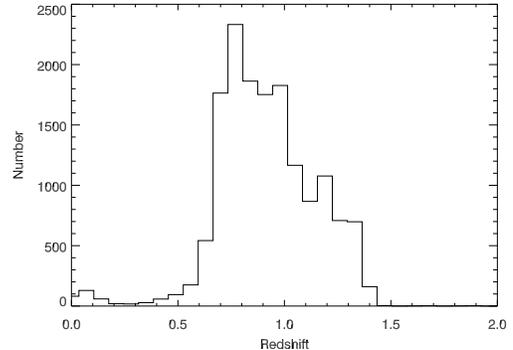}
\caption{The redshift distribution of DEEP2 galaxies in the equatorial fields.}
\label{deep2_z}
\end{figure}

This method generates a randomized galaxy-absorber pair sample, since
inclusion depends only on the relative position of the two objects.
The galaxies and QSOs will retain the selection functions of their
respective surveys, which are unrelated to our present science goals.
The quasars are being used for background illumination only.  Although
the galaxy parent catalog reflects DEEP2's selection process, we show
in Section 2.3 that it contains a mixture of spectral classes
(i.e. both star forming and early type) and hence may be used to study
the association of star formation with \mgii incidence.

We choose \mgii because it has been studied extensively at lower
redshift, and it falls directly in the optical window through DEEP2's
redshift window.  Ly $\alpha$ falls below the 3000 \AA\ atmospheric
cutoff.  \civ could be visible at the high redshift end of the sample,
but this range contains few galaxy-sightline pairs and our
observations did not yield any galaxy-absorber pairs where \civ would
be visible.  In one case we detect \ion{Fe}{2} in conjunction with
\mgii.

\subsection{Quasar Selection}

To select target quasars for observation, we cross correlated the
DEEP2 DR3 public redshift catalog\footnote{http://deep.berkeley.edu}
(containing $\sim$ 46,000 objects) with the DR6 quasar catalog
produced by the SDSS.  Magellan access necessitated restricting our
targetable declination range to the two equatorial fields in DEEP2.
We also required that $z_{QSO}>1.0$, i.e. that the quasar lie behind
much of the DEEP2 survey.  These cuts yielded a final tally of 46
Sloan quasars.

For each of these quasars, we then identified all galaxies from the
DEEP2 catalog having impact parameters of $30^{\prime\prime}$ or less
from the quasar's line of sight.  We excluded all galaxies with a
``redshift quality'' flag of 2 or lower as unreliable, following the
recommendation of the DEEP2 DR3 website.  Only galaxies with $0.6 <
z_{\rm galaxy} < z_{QSO}$ were considered for each quasar.

Figure \ref{deep2_z} shows the DEEP2 redshift histogram; in this range
the angular diameter scale subtended per arcsecond ranges from $6.7$
to $8.5$\hh kpc.  For a fiducial angular scale of 8 kpc per arcsecond,
a 100 kpc impact parameter therefore corresponds to 12.5 arcseconds on
the sky.  Ten SDSS sightlines intercept a galaxy within this radius;
these quasars form the core sample observed at Magellan.

Although not selected to do so, these same sightlines also
serendipitously pass by other galaxies at larger impact parameter.
The sample is much less complete at these larger radii, but it does
contain 5 galaxies with $100<b<200$ kpc, and many more at larger radii.

\subsection{Quasar Observation}

The ten SDSS quasars were each observed with the MagE spectrograph on
the Magellan Clay telescope.  The median QSO in our sample has
$g^\prime = 20$, with the total sample ranging from $17.95 < g^\prime
< 21.2$.  MagE is a moderate resolution UV/optical echellette that
delivers spectra from $3100$ \AA\ to 1 $\mu m$\citep{Marshall}.  A 0.7
arcsecond slit was used, corresponding to a resolution of $R=6000$, or
50 km/s.  The observations were taken under photometric conditions
during January and October of 2008.  Table 1 lists observed quasars
along with total exposure times.

\begin{deluxetable}{cccc}
\tablecolumns{4}
\tablewidth{0pc}
\tablecaption{MagE Quasar Observations}
\tablehead{
\colhead{SDSS Object} &
\colhead{$z_{QSO}$} &
\colhead{$g^\prime$} &
\colhead{Exp. Time (s)}}
\startdata
233023+0008 & 1.00 & 19.92 & 2$\times$1200\\
023101+0024 & 1.05 & 20.92 & 5$\times$1200\\
232536+0019 & 1.21 & 19.78 & 4$\times$1200\\
232634+0021 & 1.25 & 20.71 & 3$\times$1200\\
023123+0044 & 1.26 & 20.30 & 6$\times$1200\\
232632-0003 & 1.27 & 20.32 & 2$\times$1200\\
023144+0052 & 1.61 & 20.58 & 4$\times$1200\\
022639+0043 & 1.66 & 20.41 & 2$\times$1200\\
232707+0003 & 1.75 & 20.55 & 2$\times$1200\\
233050+0015 & 1.94 & 20.01 & 2$\times$1200\\
\enddata
\end{deluxetable}

Raw data from the spectrograph were processed using the MASE IDL
pipeline \citep{mase}.  MASE performs poisson-limited sky subtraction
on the 2D echelle frames using a wavelength-subsampled grid.  Object
profiles are optimally extracted while iteratively converging to a
joint model of the object and sky for each exposure.  The unbinned
extracted spectra are then combined and finally projected onto a
regular wavelength grid.  The final spectra exhibit signal-to-noise
ratios ranging from 5 to 23 with a mean of 11.6 per pixel over the
$\lambda=4500-6700$ \AA ~range relevant to the detection of \mgii in
our redshift search range.

\subsection{Galaxy Sample}
\begin{figure}
\includegraphics[scale = 0.5, clip = true, trim = 50 20 0 10]{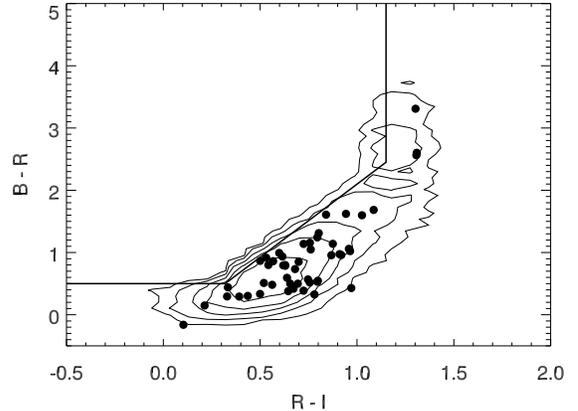}
%\plotone{f2.eps}
\caption{
Color space distribution of spectroscopically confirmed DEEP2 DR3
galaxies, shown in density contours for clarity.  Galaxies included in
our sample are shown as solid dots.  The DEEP photometric cuts define
the upper left diagonal boundary of the distribution.  Our \mgii
sample galaxies span the full range of galaxies passing the color
cuts, and are statistically consistent with a random draw from the
survey.  }
\label{deep_color}
\end{figure}

Our sample of galaxies is drawn solely from the DEEP2 survey, DR3.
From this catalog, we selected the 5 nearest galaxies for each
sightline, 50 in total.  The impact parameter of the most distant
galaxy therefore varies between fields but is no less than
$250h_{71}^{-1}$ kpc for any field.  The resulting galaxy sample spans
$0.662 \le z \le 1.409$ with a median of 0.868.  Its color
distribution is roughly uniform, containing galaxies at extremes of
both the red and blue ends of DEEP's selection region.  Figure
\ref{deep_color} shows the overall DEEP survey color distribution,
with galaxies chosen for this sample highlighted.  The color
distribution of \mgii candidate galaxies is consistent with a random
draw from DEEP2, in the group passing the color cuts.  To confirm
this, we ran a 2-sample, 2-dimensional Kolmogorov-Smirnov (KS) test on
the distributions of $(R-I)$ and $(B-R)$ colors.  The KS test admits
the possibility (i.e. rules out at only the $P=25$\% level) that our
sample and the full DEEP2 $0.6<z<1.4$ sample are drawn from the same
color space.  Likewise a one-dimensional two sample KS test on the
redshift distribution fails to distinguish between our sample and that
of DEEP2 in our redshift range ($P=30$\%).

We have excluded DEEP2 objects in the Extended Groth Strip (EGS),
where different selection criteria were used; we do not observe any
QSOs in the EGS because of its northern declination and so have no
\mgii candidates within this group.

\begin{figure}
\includegraphics[scale = .45, clip = true, trim = 10 0 0 0]{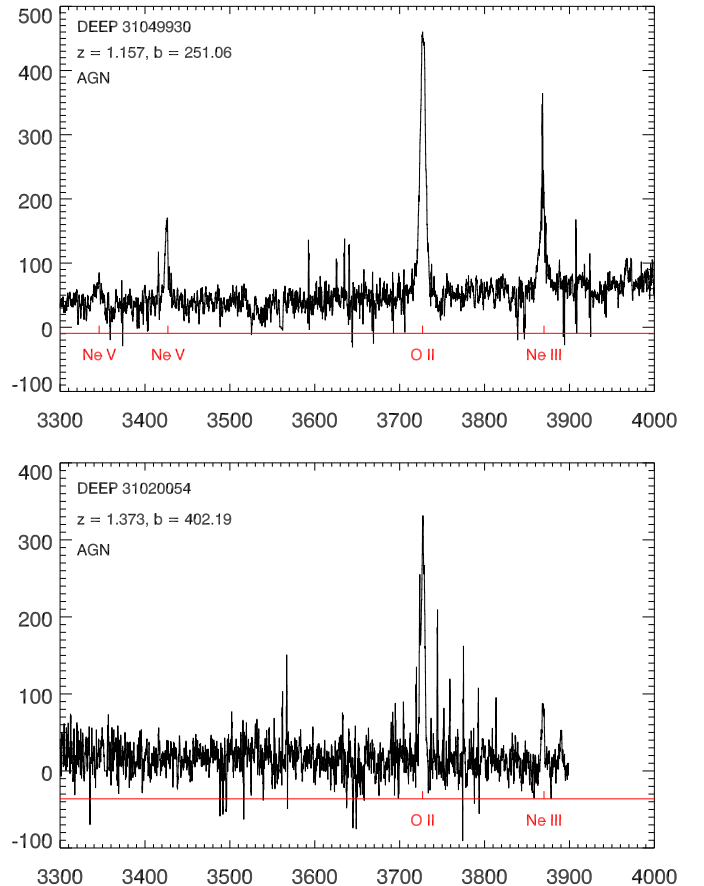}
%\plotone{f3.eps}
\caption{Spectra of AGNs in sample.  31049930 displays a broadened
  \oii peak, strong \ion{Ne}{3}, and strong \ion{Ne}{5}.  31020054
  displays a broadened \oii doublet and \ion{Ne}{3} peak.}
\label{agn_spectra}
\end{figure}

The 50 galaxy spectra extracted from the DEEP archive were inspected
in order to search for AGN signatures, and two candidates were
identified: one definite (31049930), which displays strong
\ion{Ne}{3} and \ion{Ne}{5} lines and a broad \oii peak; and one very
likely (31020054), which displays strong \ion{Ne}{3} lines and a
broad \oii peak.  These spectra can be seen in Figure
\ref{agn_spectra}.  This gives an AGN fraction of 4\% of the
population.  Neither of the AGN candidates are located within the
primary $b \le 100$ kpc sample.

The DEEP2 survey utilizes an apparent magnitude cut of $B<24.1$, which
at our median redshift corresponds to $M_B -5\log(h)= -19.6$, ignoring
$k$-corrections (we account for these on a per-galaxy basis below).
At the extreme low and high redshift ends of the sample, the limiting
absolute magnitude changes by approximately $\pm 1$.  For consistency
with low redshift studies we have used $M_B^*=-19.8+5\log(h)\approx
-20.5$, so the magnitude limit (at z=0.87) corresponds to roughly
$\sim 0.8L^*$ on this scale.  Several galaxies are slightly fainter
than this on account of their lower redshift, but the bulk of our
sample consists of galaxies with $L\sim 1-3L^*$.

\section{Results}
\subsection{Searching QSO Spectra}
After the raw frames were processed using MASE, the resulting ten
spectra were inspected visually for \mgii absorption doublets at
redshifts corresponding to the galaxies associated with each line of
sight.  Where a doublet was detected, $W_r$ was measured directly from
the spectral pixels.  Where no doublet was detected, a Gaussian
absorption profile was fitted to the data at the redshift of the
associated galaxy to estimate a $3\sigma$ upper limit.  A total of 5
doublets were detected at a $z$ coincident with a DEEP galaxy; for 2
sightlines a pair of galaxies were identified near the absorber's
redshift, for a total of 7 associated DEEP galaxies.  Five \mgii
doublets were detected within the DEEP redshift range but with no
corresponding DEEP galaxy, and six additional \mgii systems were
foundoutside of the DEEP2 range.  All \mgii detections within the
DEEP2 redshift window are listed in Table 2.  DEEP objects with no
corresponding \mgii detections are listed along with upper limits on
their \mgii equivalent width in Table 3.

\subsection{Measuring SFR}
\label{sfr}
Since many previous papers have examined the connection between \mgii
absorption, galactic winds, and star formation rate (SFR), we measure
the SFR of galaxies in our sample.  Although the preferred indicator
of star formation rate (SFR) is the \ha line, this wavelength is
redshifted out of all DEEP spectra in our sample.  We therefore
measure SFR using the emission line luminosity correlations
empirically derived in \citet{moustakassfr2006}, which give SFR as a
function of $B$-band luminosity and either \oii or \hbeta line
luminosity.  We preferentially use \hbeta where available, since it is
a more reliable tracer less subject to dust attenuation.

\begin{figure}
%\epsscale{1.2}
%\plotone{f4.eps}
\includegraphics[scale=0.48, clip = false, trim = 10 10 0 0]{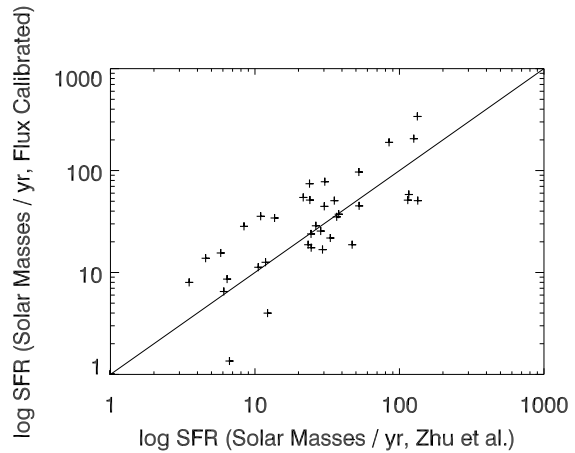}
\caption{Comparison of measurements of the star formation rate using
  two different methods described in the text.  }
\label{sfr_compare}
\end{figure}

The DEEP spectra are not flux-calibrated so line luminosities cannot
be calculated in the normal fashion.  We therefore tested two separate
methods for determining the SFR from DEEP spectra, and compared for
consistency.  The first method utilizes an archival flux calibration
of the DEIMOS spectrograph, as documented in \citet{lemaux} and on the
DEIMOS website\footnote{http://www.ucolick.org/$\sim$ripisc/results.html}.
This method has the advantage of producing direct \oii line fluxes and
luminosities, but, lacking spectrophotometric standard observations,
it is not a true flux calibration and is subject to time-dependent
variation in weather and instrument conditions.  We use this as our
primary SFR measure throughout the paper.

The second method, described in \citet{zhuoii2009}, combines un-fluxed
spectra with calibrated photometry to bootstrap an estimate of the
\oii line luniosity.  First we measure the equivalent width of the
line using an estimated average continuum.  We then use {\tt
  kcorrect}\footnote{v.4\_2, \citet{kcorrect}} to reconstruct a
galactic SED using the DEEP2 broadband photometry.  We take the
reconstructed model continuum value at the line, and multiply it by
the equivalent width in order to recover the line flux.  This flux is
then used to calculate the line luminosity and thus SFR.  In cases
where no emission is detected at the line location, we set the
SFR to zero after inspecting the line to ensure it is continuum, since
the SFR correlations do not handle nondetections well and produce
unreasonably large upper limits.

Figure \ref{sfr_compare} shows results from both procedures.  In
general the two methods agree fairly well, having random scatter of
$\sim 0.3-0.5$ dex but no systematic offsets.  Our SFR values ranged
between 0.5 $\le$ log(SFR [M$_{\odot} $ yr$^{-1}]) \le 2.0$.  AGN are
automatically assigned 0 SFR since emission from the AGN renders
emission line SFR diagnostics inapplicable.  Rest-frame $M_B$ values
are also calculated using {\tt kcorrect}; the SFR values and absolute
magnitudes are listed in Tables 2 through 5.

\subsection{Notes on Individual Systems}
Here we describe properties of the seven sample galaxies hosting \mgii
absorption, indicated in Figures 5 through 9.  For comparison
purposes, we also include plots of the six lowest-$b$ galaxies that
had no corresponding \mgii detections in Figure 10.  Full spectra of
these galaxies are presented in Figure 11 (detections) and Figure 12
(non-detections).  All detections have $W_r \le$ 0.8 \AA, and all
non-detections have 3$\sigma$ upper limits of $W_{lim} \sim 0.3$ \AA.
The measured properties of each positively identified galaxy-absorber
pair are reported in Table 2.  Plots of galaxy spectra are shown in
the galaxy's rest frame.

\subsubsection{SDSS232634+0021, DEEP 31052516}

\begin{figure}
\includegraphics[scale = 0.5, trim = 15 120 20 50, clip=true]{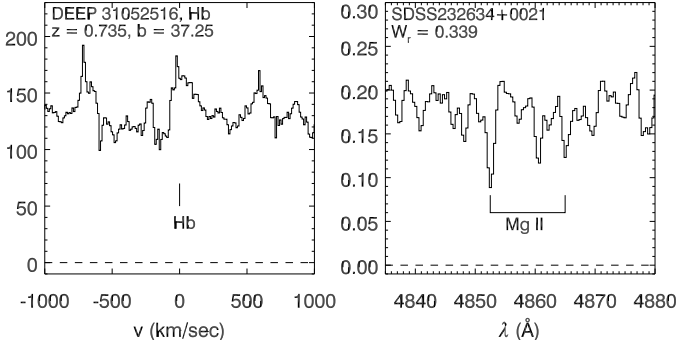}
%\epsscale{1.0}
%\includegraphics{f4.eps}
\caption{Detail of the $z=0.735$ galaxy-absorber pair. }
%\epsscale{1.0}
\end{figure}

\textbf{$z_{gal}$ = 0.7346, $b$ = 37.25 kpc, $W_r$ = 0.339 \AA\ +/-
  0.067, $\Delta v$ = -266 km/s} This galaxy's redshift is low enough
that the \hbeta line is accessible; however, there is no detected
emission at the appropriate wavelength.  Rather, we see strong Ca H+K
absorption and a 4000 \AA\ break.  This galaxy shows little to no
signs of star formation, but with a rest-frame $M_B$ of -21.21, it
represents the brightest galaxy in the detected sample; this combined
with its spectral features classify it as a luminous red galaxy (LRG).
However, despite its low impact parameter (the lowest in the sample)
and high intrinsic brightness, its absorption feature is weak.
\subsubsection{SDSS233050+0015, DEEP 32024543}

\begin{figure}
\epsscale{0.8} 
%\plotone{f5a.eps}
%\plotone{f5b.eps}
\includegraphics[scale=0.5,trim = 15 120 10 30,clip=true]{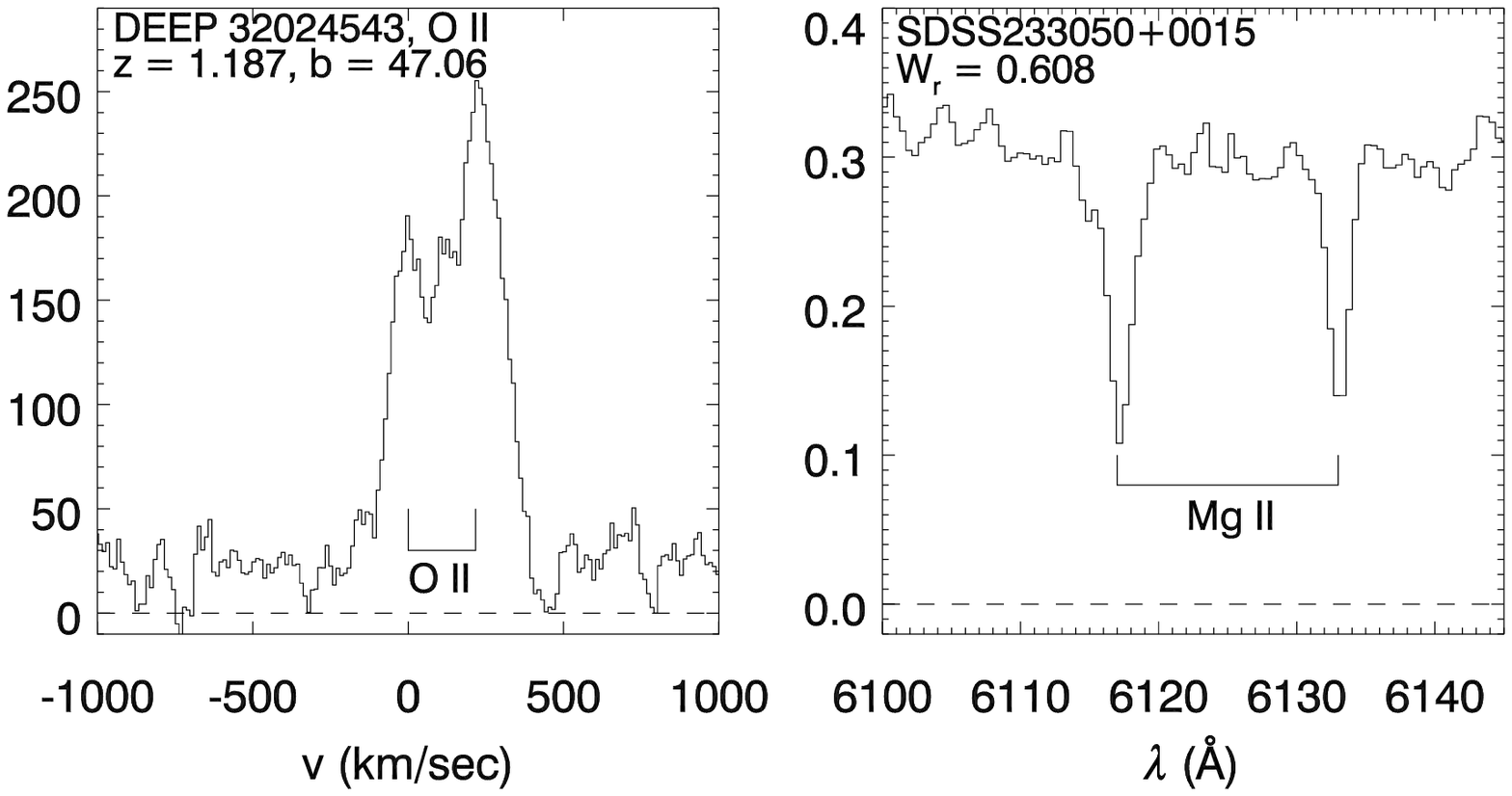} 
\includegraphics[scale=0.5,trim = 15 120 10 30,clip=true]{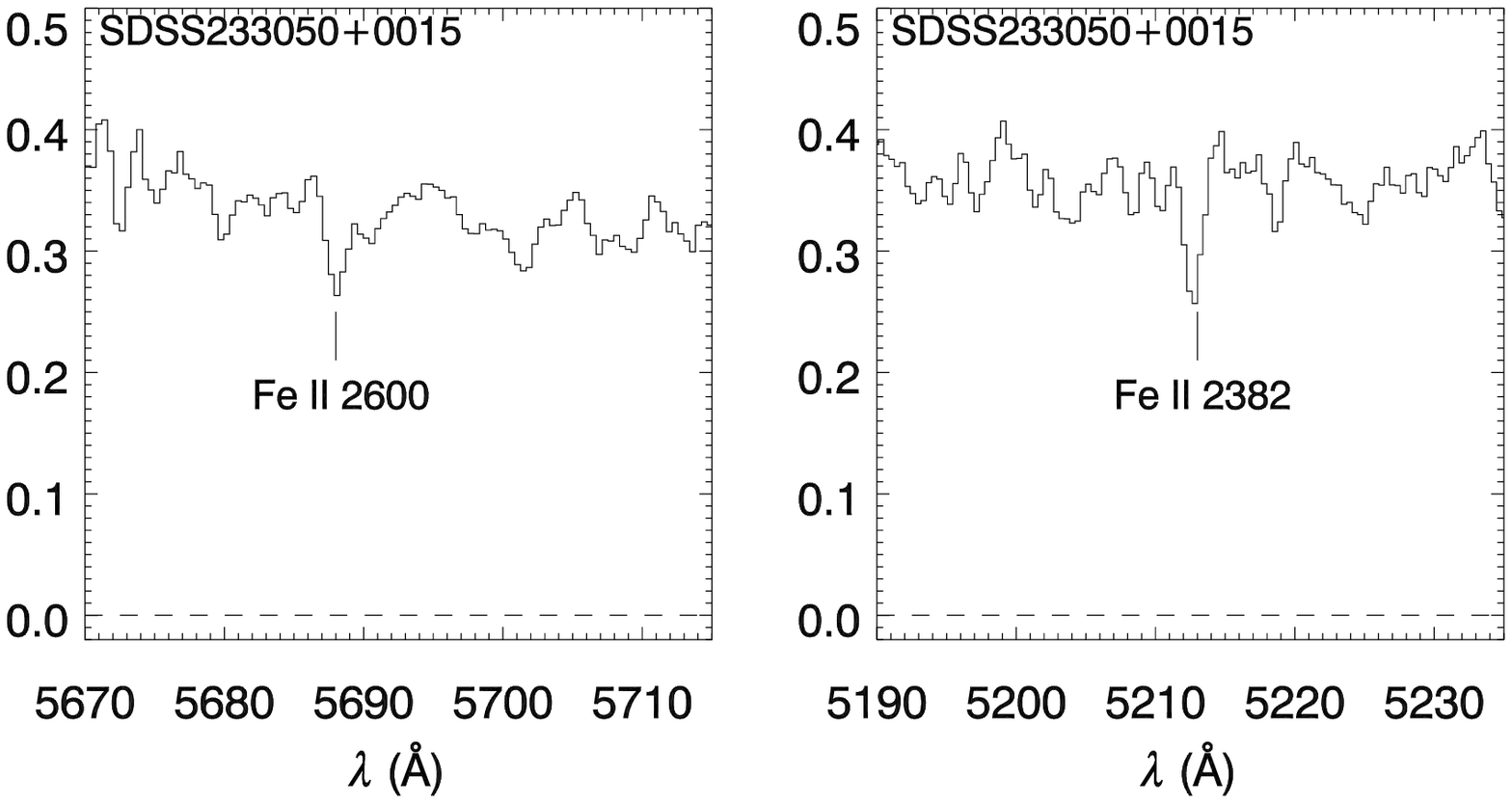}
\caption{Detail of the $z=1.187$ galaxy-absorber pair. }
\end{figure}

\textbf{$z_{gal}$ = 1.1875, $b$ = 47.06 kpc, $W_r$ = 0.608 \AA\ +/-
  0.048, $\Delta v$ = -95 km/s} DEEP 32024543 displays a strong \oii
peak, indicative of strong star formation.  Its corresponding \mgii
absorption feature is strong and narrow, and two additional
\ion{Fe}{2}\ lines are visible.  The \ion{Fe}{2} 2600 line has
$W_r=192\pm 52$ m\AA, while the 2382 line has $W_r = 129\pm 46$ m\AA.
\subsubsection{SDSS233023+0008, DEEP 32014809}

\begin{figure}
\epsscale{1.2}
\includegraphics[scale = 0.5, trim = 15 120 10 50, clip=true]{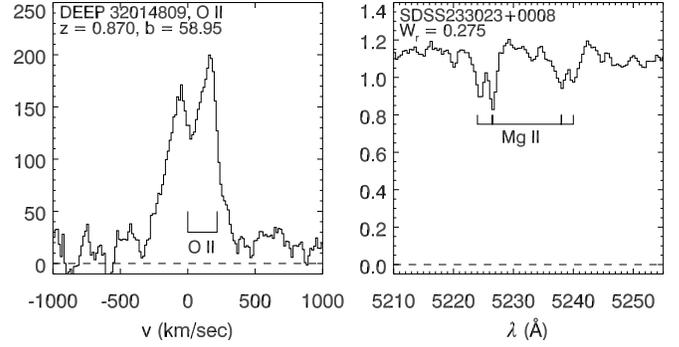}
%\plotone{f6.eps}
\caption{Detail of the $z=0.870$ galaxy absorber pair. }
\end{figure}

\textbf{$z_{gal}$ = 0.8702, $b$ = 58.95 kpc, $W_r$ = 0.275 \AA\ +/-
  0.030, $\Delta v$ = 420 km/s} DEEP 32014809's \mgii absorption line
displays an interesting velocity structure, showing not one but two
distinct doublets at a corresponding dispersion velocity of $\sim$50
km/s, indicating interceptions by two separate clouds of \mgii moving
with individual velocities.  The galaxy itself displays a strong \oii
peak indicative of strong star formation.

\subsubsection{SDSS232536+0019, DEEP 31050080, 31050146}

\begin{figure}
\includegraphics[scale = 0.5, trim = 15 120 10 50, clip=true]{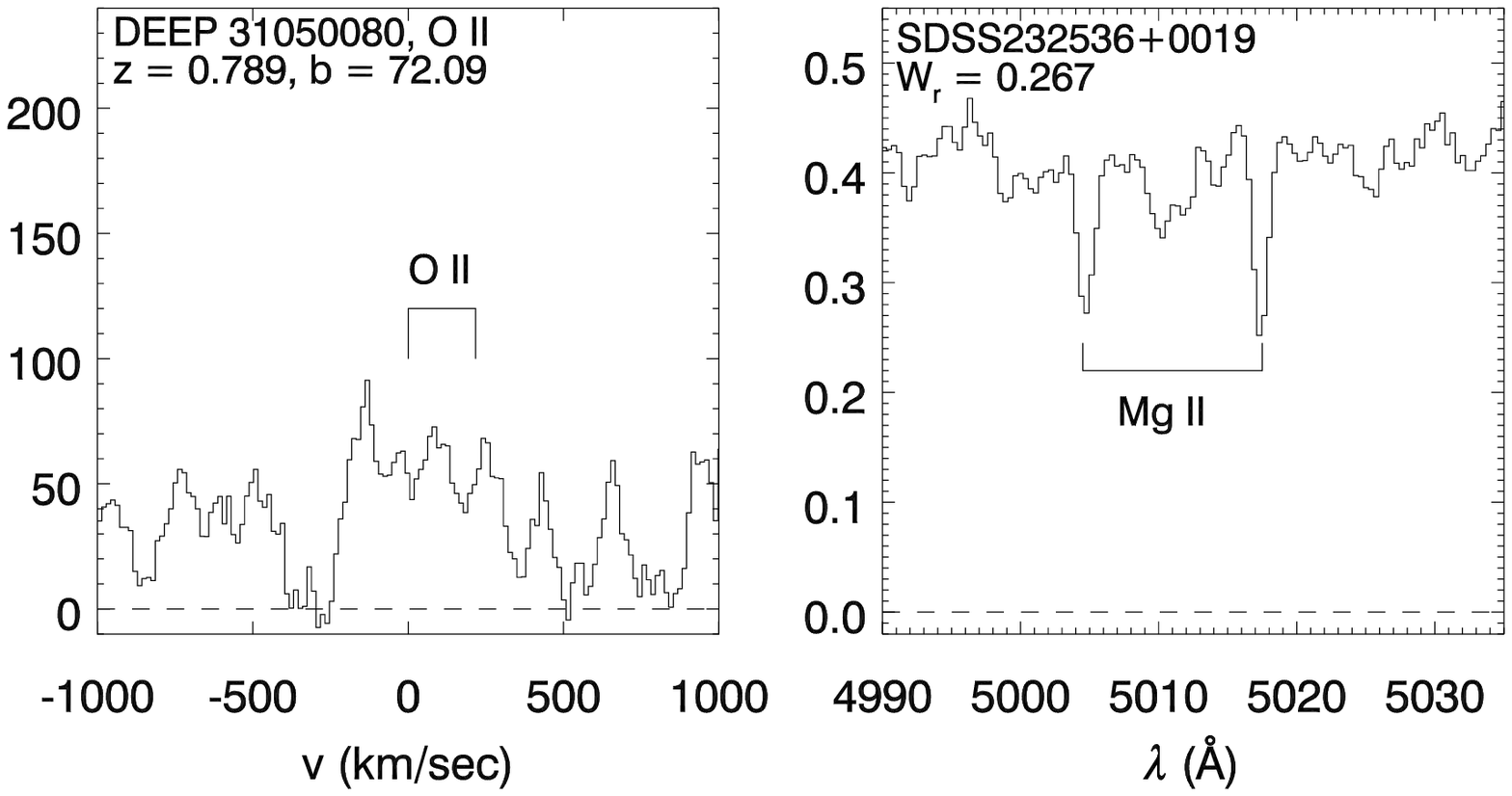}
\includegraphics[scale = 0.5, trim = 15 120 10 50, clip=true]{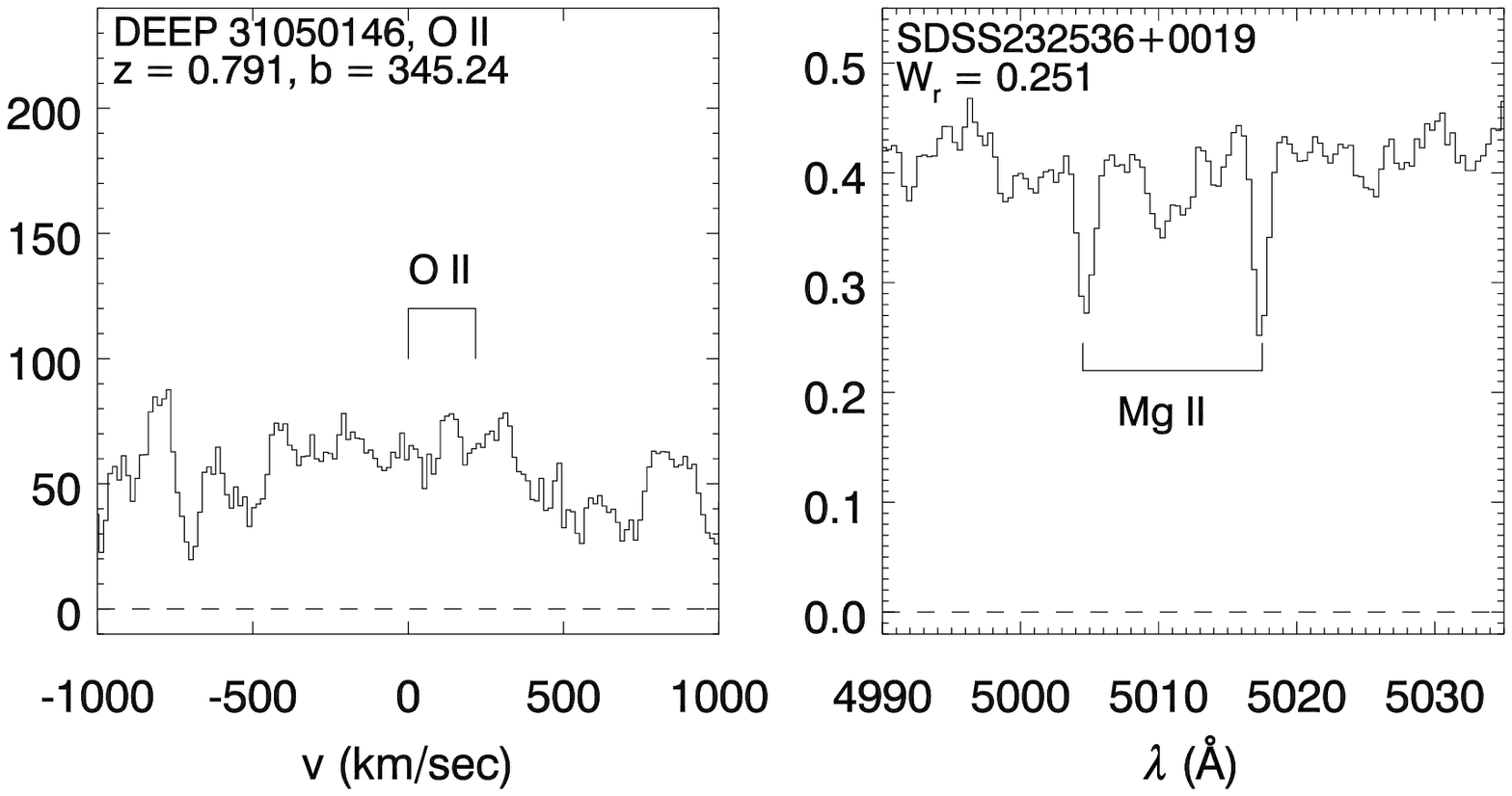}
%\epsscale{1.15}
%\includegraphics[scale=0.8,trim = 15 120 10 30, clip=true]{f8a.eps}
%\includegraphics[scale=0.8,trim = 15 120 10 30, clip=true]{f8b.eps}
\caption{Detail of the $z=0.790$ galaxy absorber pair. }
\end{figure}

\textbf{$z_{gal}$ = 0.7894/0.7912, $b$ = 72.09/345.24 kpc, $W_r$ =
  0.263 \AA\ +/- 0.040, $\Delta v$ = 387/-151 km/s} These two galaxies
have very similar redshifts and both exhibit spectral features typical
of early-type galaxies, lacking emission at the wavelength of the \oii
doublet and displaying strong 4000 \AA\ breaks; these objects may be
members of a galaxy group.  Although one of these galaxies is at a
much lower impact parameter than the other, the galaxies visually
bracket the QSO on the sky and the \mgii absorption line is located at
a redshift between the two.  The absorbing gas may therefore be
associated with the group rather than either galaxy specifically.
This absorption line also displays an odd asymmetry in the doublet
structure, with the normal ratio of \mgii line depths reversed.  This
is not an artifact of stacking a bad frame during the reduction
process, as the same asymmetry appears in each individual exposure.
We examined alternative identifications for this system besides \mgii
while enforcing that $0 < z_{abs}<z_{QSO}$, which constrains the rest
wavelength to be $2266 < \lambda_{rest} < 5000$ \AA ~($z_{QSO}=1.21$).
Besides \mgii, the strongest typical QSO absorption lines in this
range are from \feii, which is ruled out based on the non-detection of
other multiplet transitions, and Ca II, which is likewise ruled out as
a doublet (and would be accompanied by strong DLA-like absorption,
which is absent here).  Given the exact match in wavelength separation
between the components, we suspect the reversed doublet ratio to be
the result of contamination by a lower redshift interloper blending
with the 2803 \AA ~line.

\subsubsection{SDSS023144+0052, DEEP 43056587, 43056310}

\begin{figure}
%\epsscale{1.15}
\includegraphics[scale=0.5,trim = 15 120 10 30, clip=true]{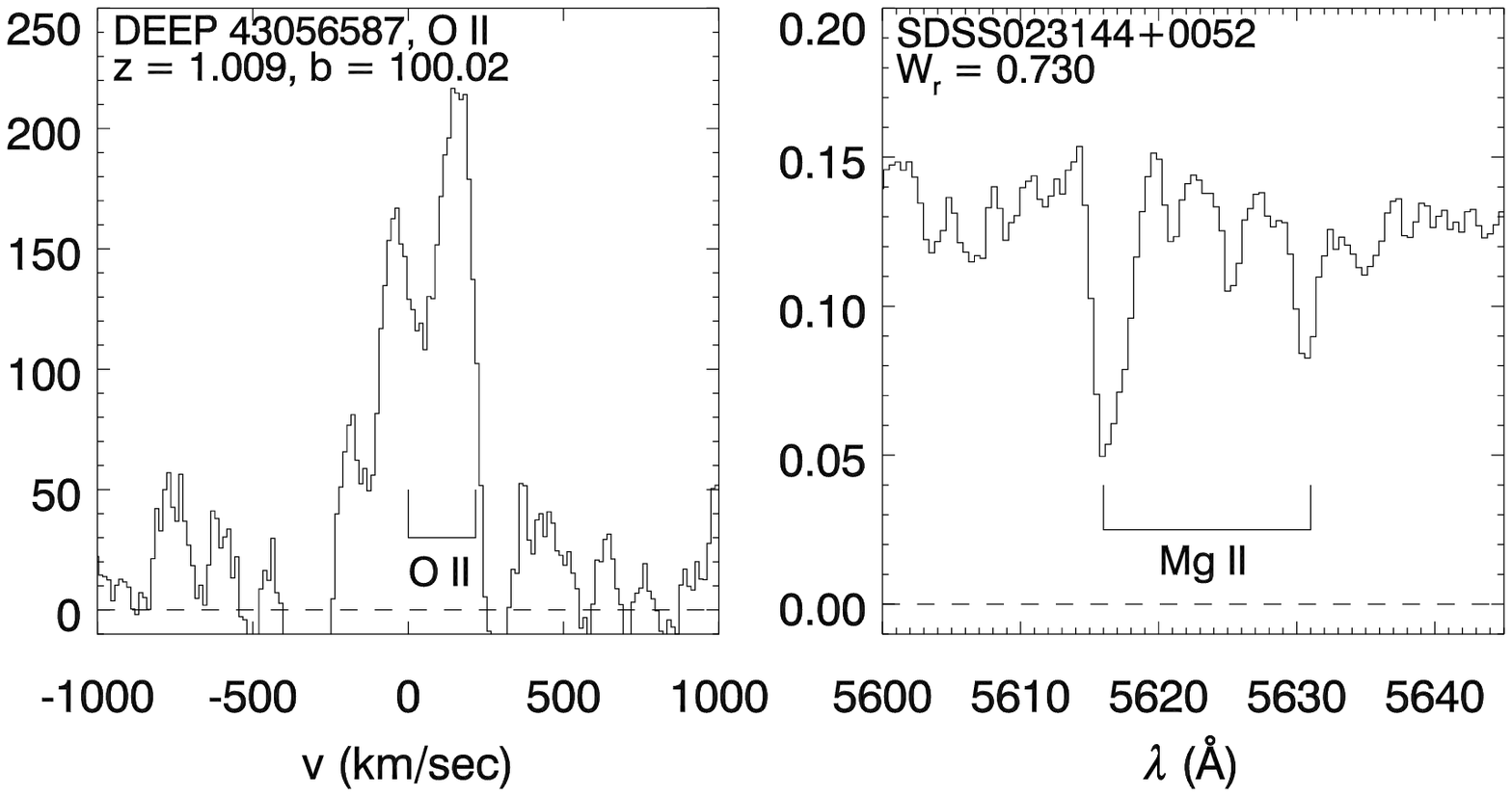}
\includegraphics[scale=0.5,trim = 15 120 10 30, clip=true]{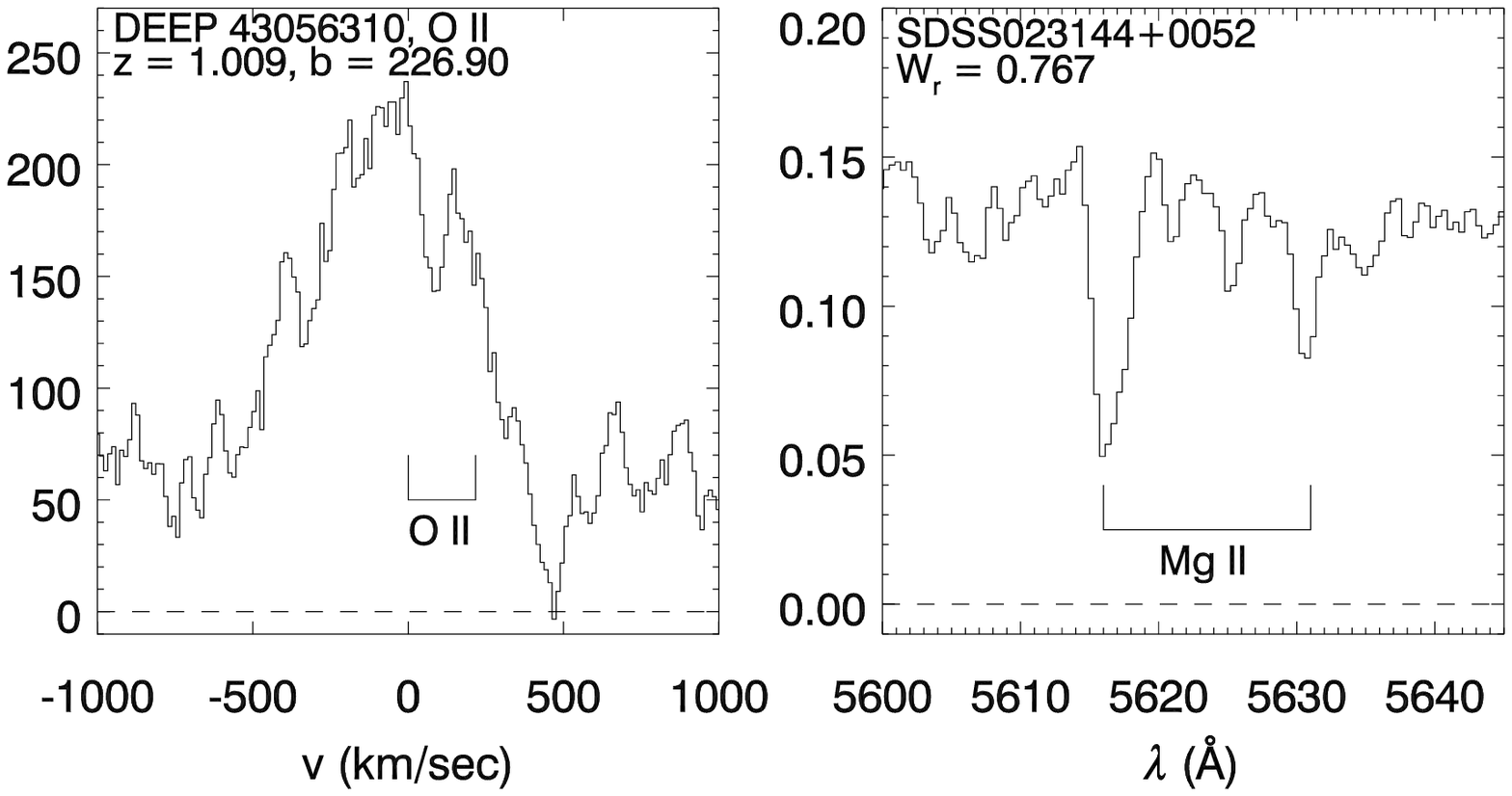}
%\plotone{mg_43056587.eps}
%\plotone{mg_43056310.eps}
\caption{Detail of the $z=1.009$ galaxy absorber pair. }
\end{figure}

\textbf{$z_{gal}$ = 1.0086/1.0091, $b$ = 100.02/226.90 kpc, $W_r$ = 0.759 \AA\ +/- 0.067, $\Delta v$ = 11/164 km/s} Although only a single
strong \mgii absorption doublet is seen in the QSO spectrum, two DEEP
galaxies are located at the corresponding redshift.  DEEP 43056587
shows a strong \oii peak, while DEEP 43056310 displays a broadened
\oii doublet, but no other characteristic AGN emission lines (e.g. of
Ne).

\begin{figure}
\includegraphics[scale = 0.4, trim = 15 120 10 50, clip=true]{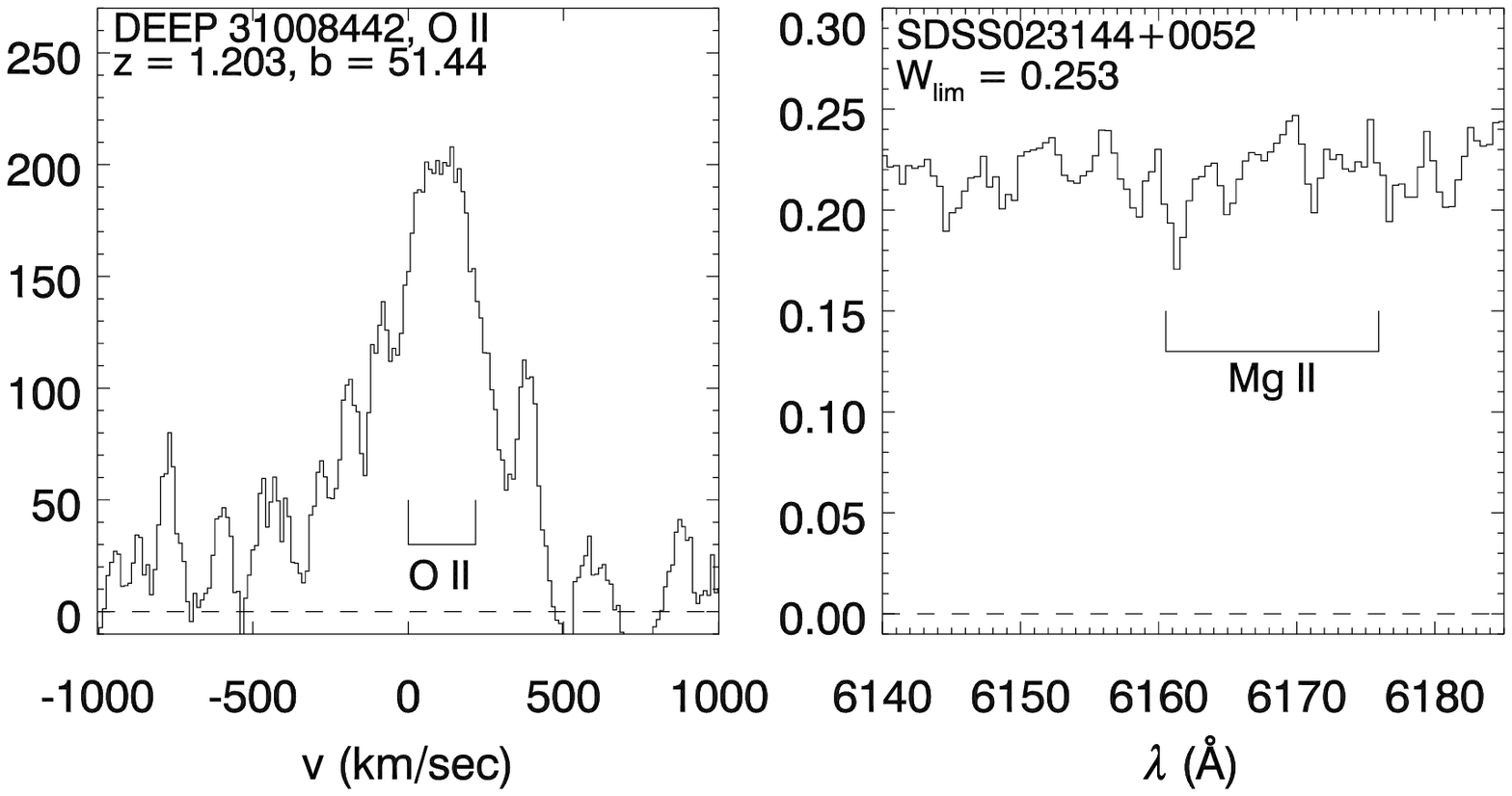}
\includegraphics[scale = 0.4, trim = 15 120 10 50, clip=true]{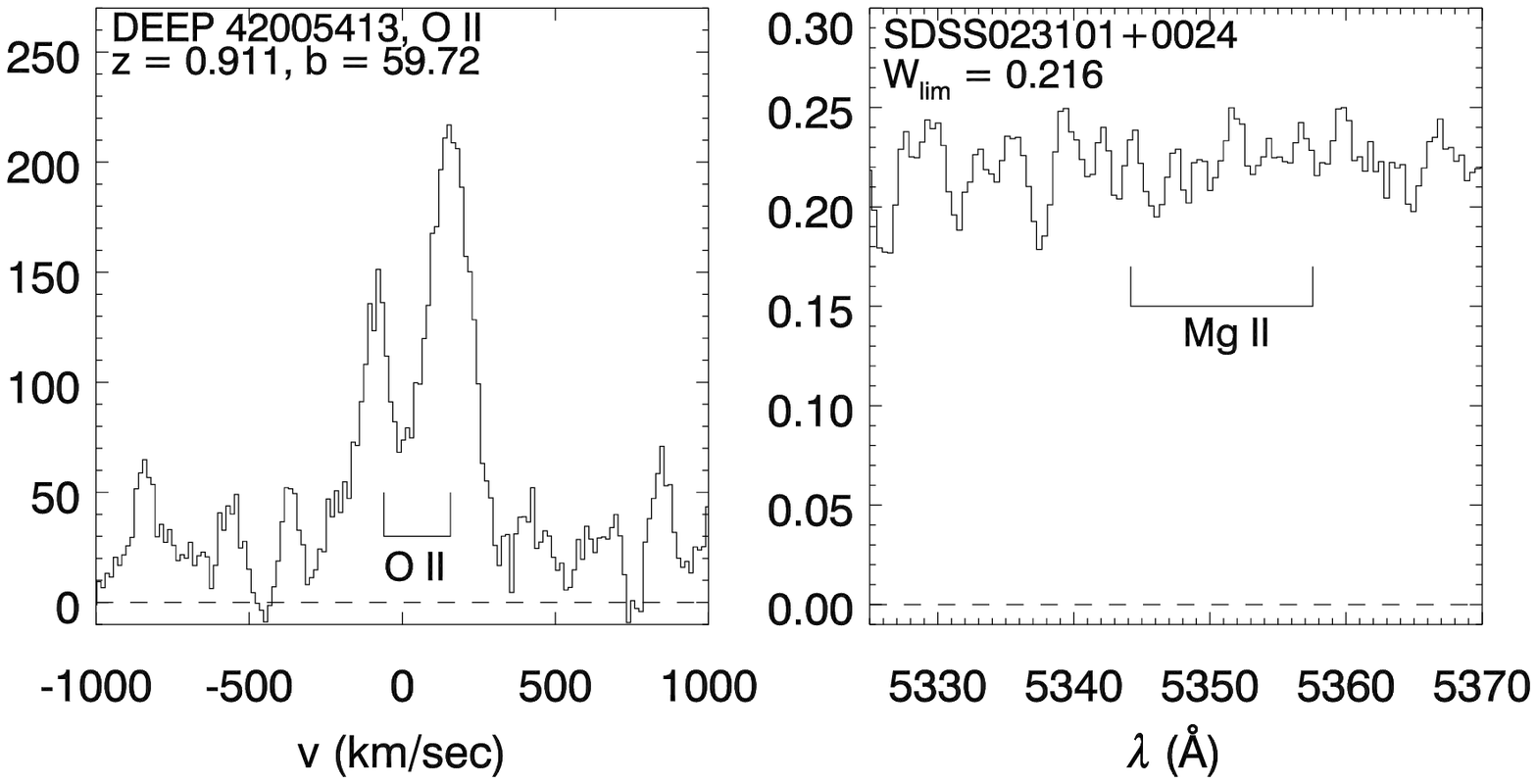}
\includegraphics[scale = 0.4, trim = 15 120 10 50, clip=true]{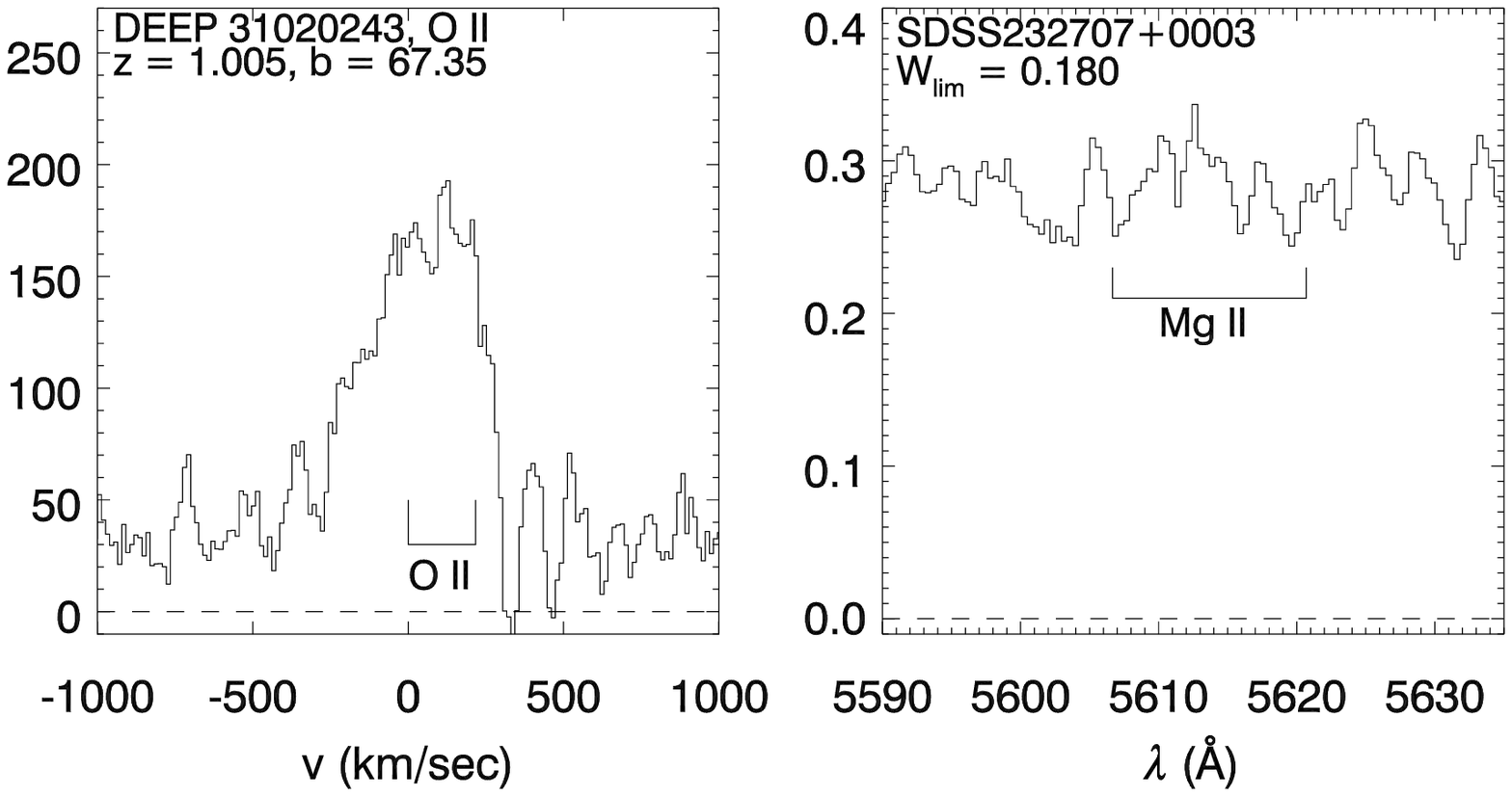}
\includegraphics[scale = 0.4, trim = 15 120 10 50, clip=true]{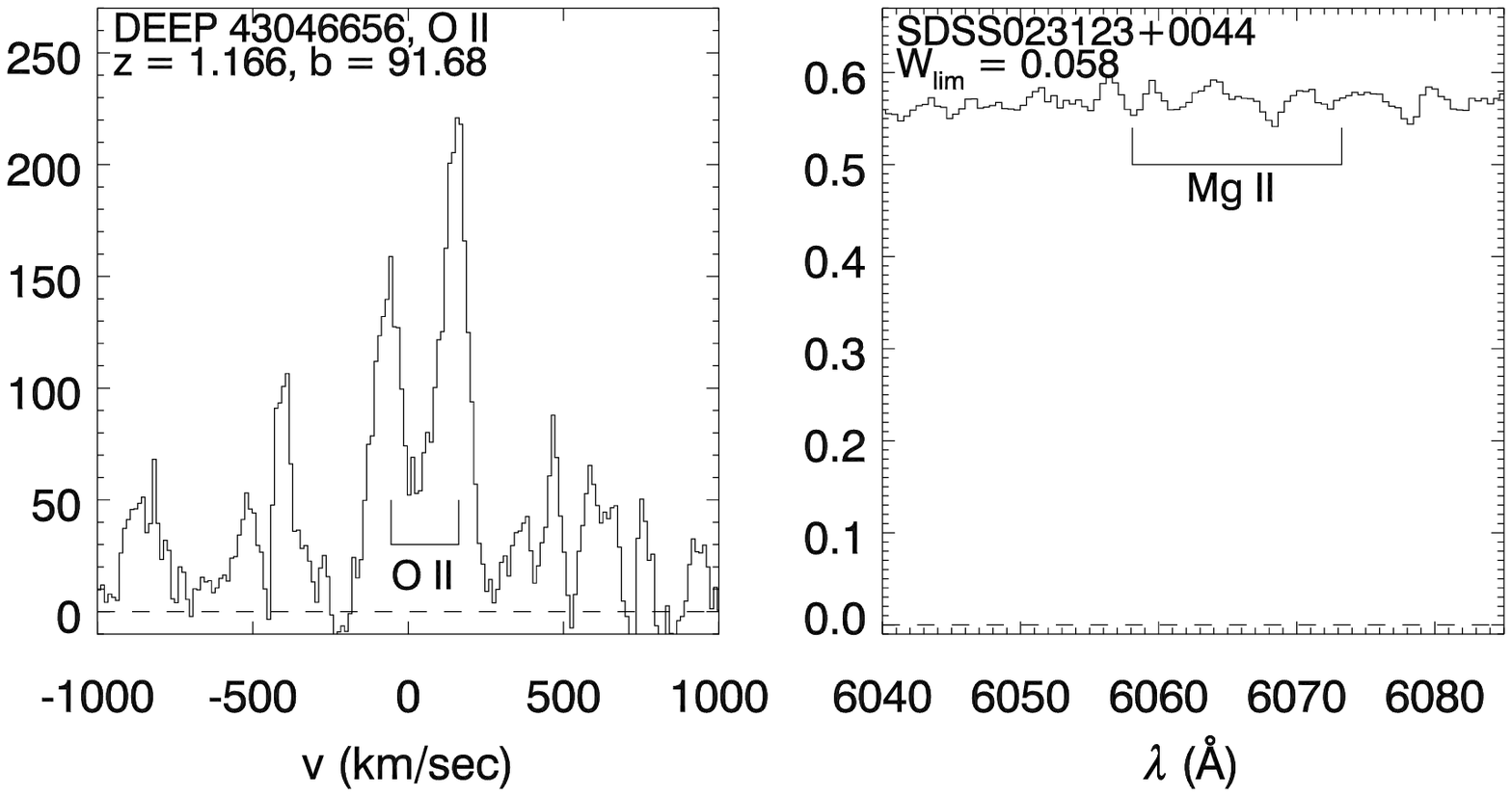}
\includegraphics[scale = 0.4, trim = 15 120 10 50, clip=true]{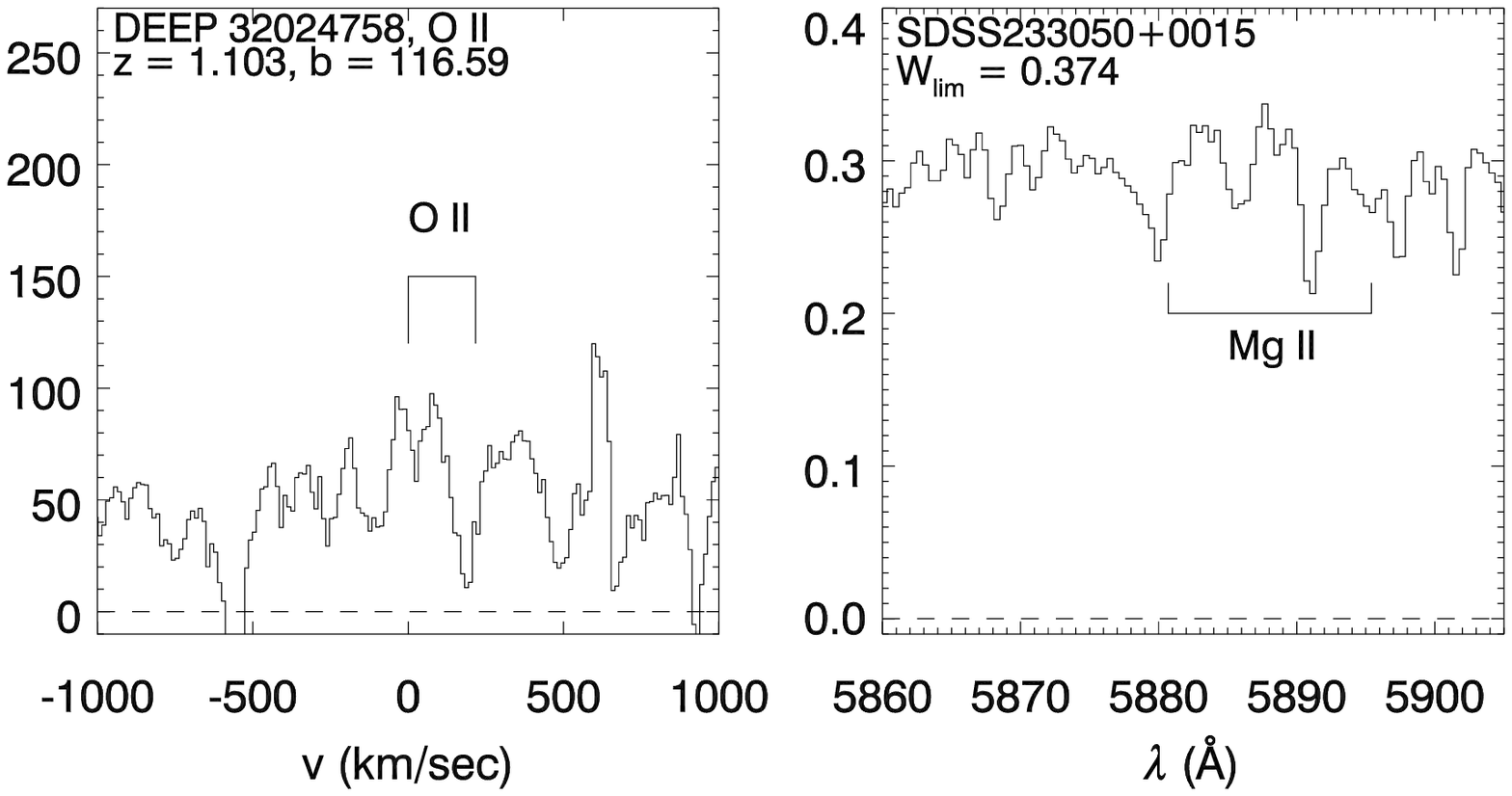}
\includegraphics[scale = 0.4, trim = 15 120 10 50, clip=true]{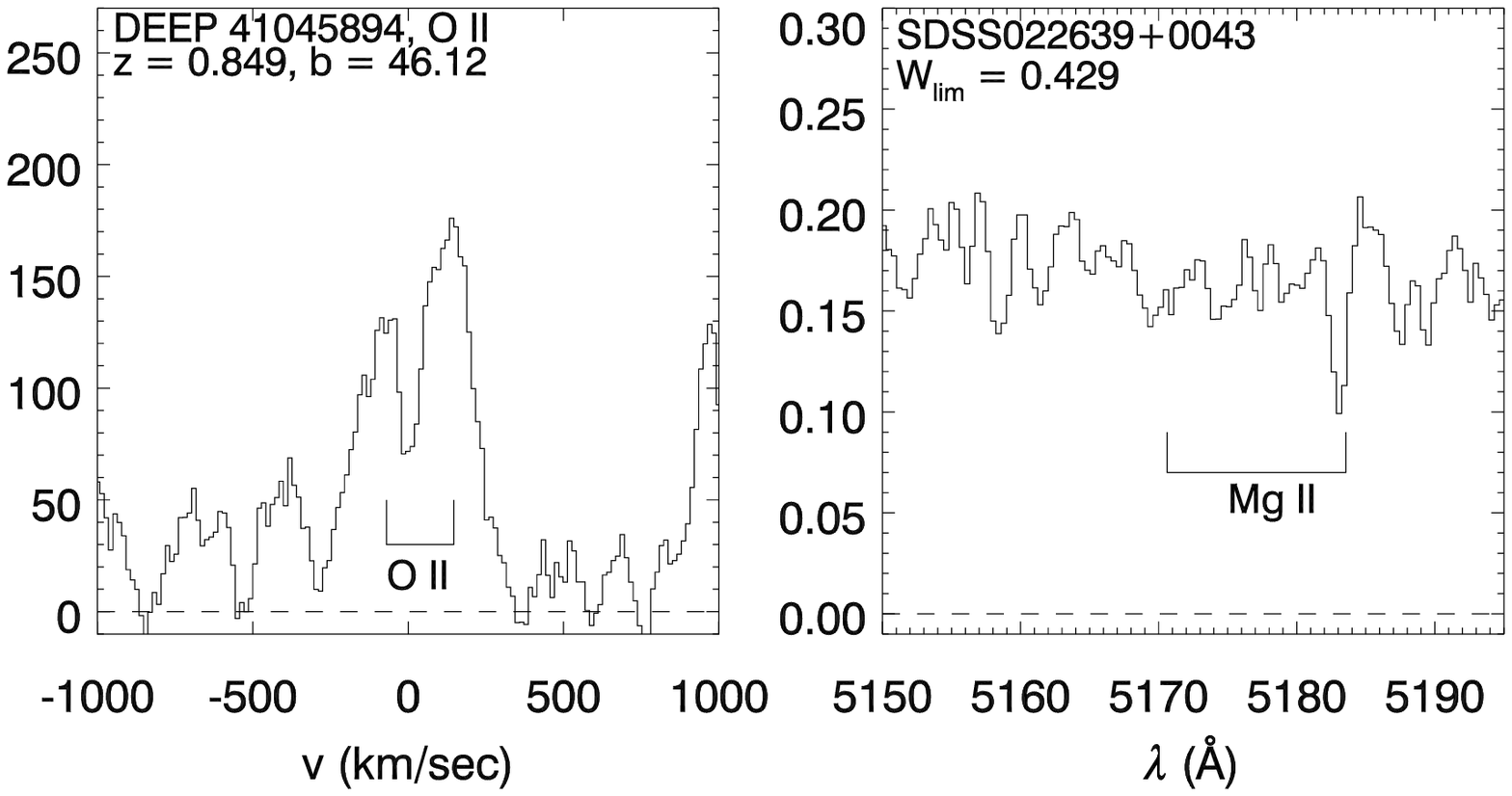}
%\plotone{mg_31008442.eps}
%\plotone{mg_42005413.eps}
%\plotone{mg_31020243.eps}
%\plotone{mg_43046656.eps}
%\plotone{mg_32024758.eps}
%\plotone{mg_31020310.eps}
\caption{Montage of six DEEP2 galaxies with $b\lesssim 100$ kpc, but
  no detected \mgii.  In each case, the left panel shows the [O II]
  emission line (or $H\beta$ for low redshift), while the right
  panel displays the \mgii region and associated $3\sigma$ upper limit
  on $W_r$.}
\label{fig:upperlimits}
\end{figure}

\begin{figure*}
%\epsscale{1.0}
%\plotone{f18.eps}
%\plotone{f11.eps}
\includegraphics[trim = 20 0 20 0, clip=true, scale = .85]{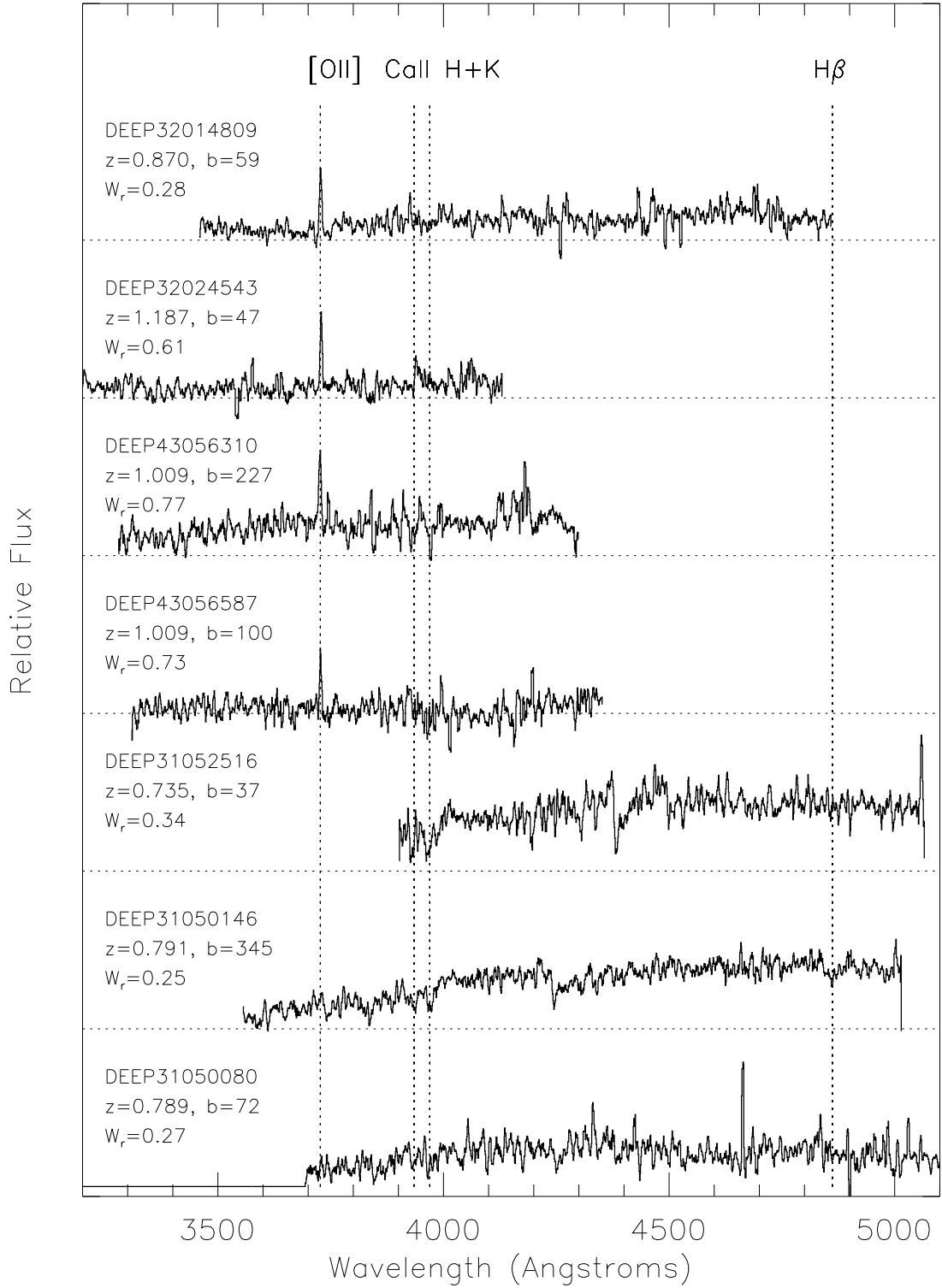}
\caption{DEEP/DEIMOS archival spectra of galaxies associated with
  \mgii absorption seen in background QSOs.  The bottom three galaxies
  display little evidence of star formation, while the bottom four
  show moderate activity.}
\label{detected_spectra}
\end{figure*}

\begin{figure*}
%\epsscale{1.0}
%\plotone{f12.eps}
\includegraphics[trim = 20 0 20 0, clip=true, scale = .85]{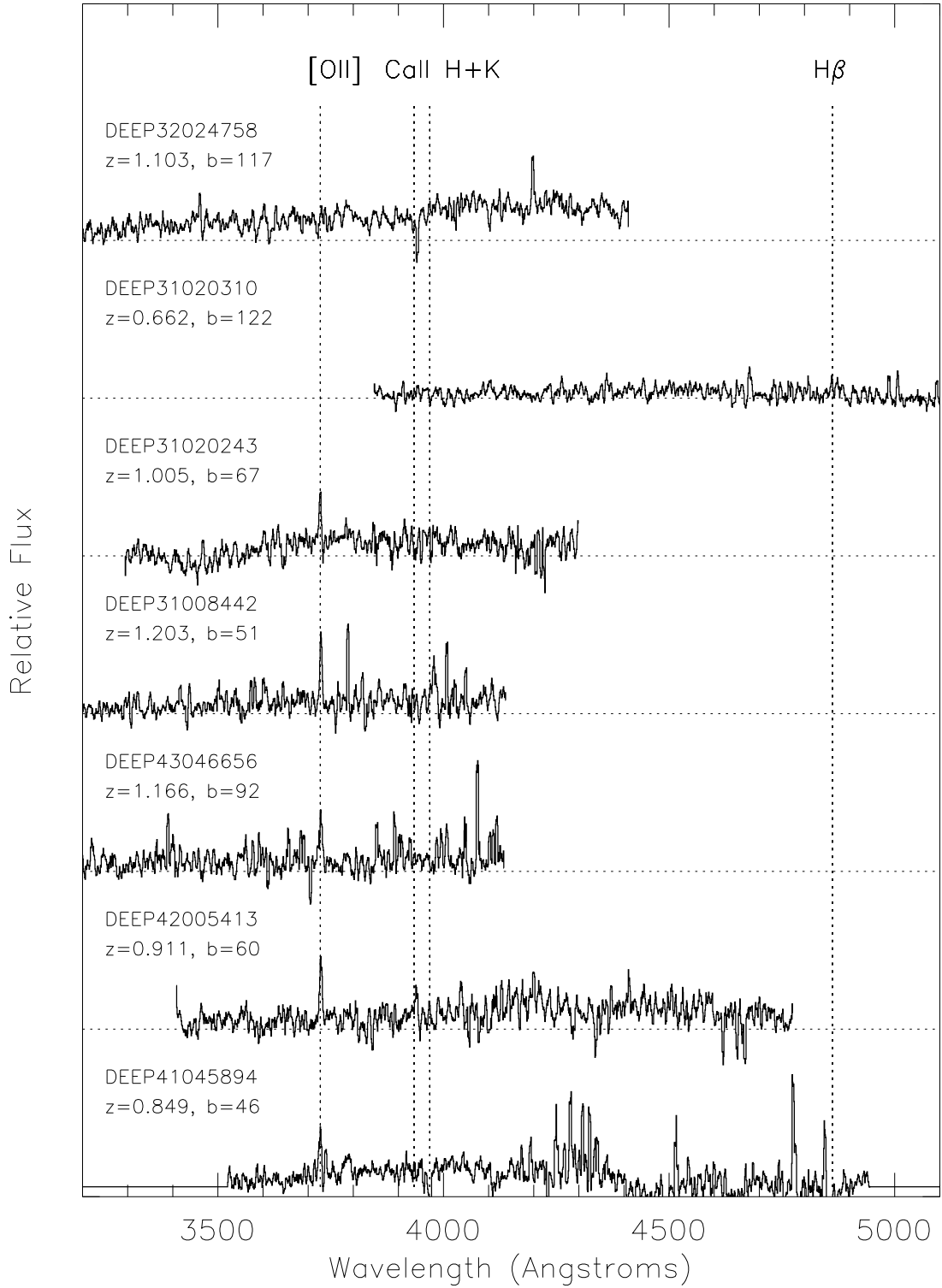}
\caption{DEEP spectra of seven galaxies for which no \mgii was
  detected in absorption.  They are chosen to have the smallest $b$ of
  the non-absorbing systems, with five of seven at $b < 100$ kpc.}
\label{nondetected_spectra}
\end{figure*}

\FloatBarrier

\subsection{Trends in \mgii Absorption}

\subsubsection{Impact Parameter}

\begin{figure}
%\epsscale{0.8}
%\plotone{f13.eps}
%\includegraphics[scale = 0.45, clip = false]{b_vs_mg.eps}
\includegraphics[scale = 0.5, clip = true, trim = 50 110 0 45]{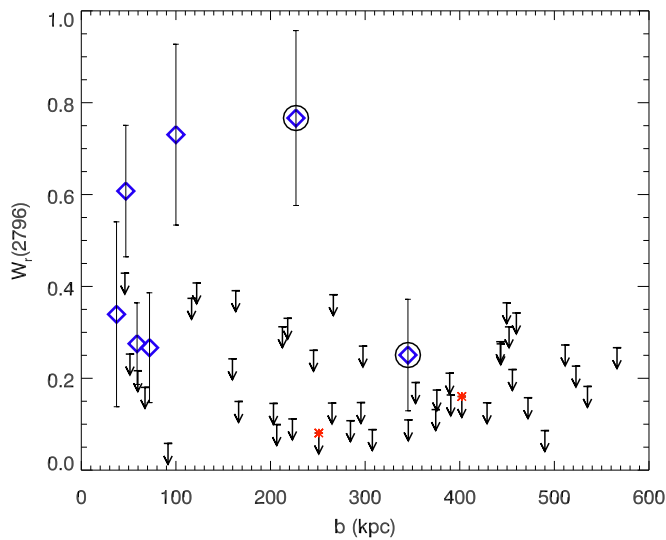}
\caption{\mgii EW vs. impact parameter $b$.  The open diamonds (blue
  in the color figure) represent detected \mgii absorption.  The
  arrows (black) represent upper limits on possible \mgii absorption.
  Black circles mark the outer members of galaxy pairs discovered at
  the absorber redshift.  Asterisks (red) mark AGNs.  See the
  electronic edition for a color version of this figure.}
\label{b_vs_mg}
\end{figure}

In Figure \ref{b_vs_mg}, we plot the \mgii equivalent widths of our
systems against their respective galaxy/QSO impact parameter.  The
bulk of \mgii detections occur in galaxies with $b\le 100$ kpc.
Beyond this limit, we find no unique galaxy-absorber pairs.  In two
sightlines we detect absorption at a redshift coincident with a galaxy
at large separation; however, in each case there is a second galaxy at
$b<100$ kpc also located at or near the same redshift. Other than this
general $b$ limit, no other strong trend emerges.

Simple counting yields a raw covering factor of $\kappa=50\%$ (5
detections vs. 5 non-detections) at impact parameters $\lesssim$100
kpc, with $\kappa=0$\% at $b>100$ kpc (discounting the double-galaxy
systems).  At first glance this matches well with the uncorrected
covering fraction of 57\% from \citet{chenandtinker2008} or 66\% from
\citet{chen2010lowz} (Hereafter CT08 and CT10).  However, both CT08
and CT10 follow the standard practice \citep{steidel_scaling,
  mgii_rotation} of renormalizing each pair's impact parameter to
account for the fact that larger galaxies should have more extended
gas haloes.  This is expressed mathematically as an empirical power
law scaling between halo radius $R_{gas}$ and galaxy luminosity $L_B$:
\begin{equation}
R_{gas}=R_{gas}^*\left({{L_B}\over{L_B^*}}\right)^\beta
\end{equation}
where $R_{gas}^*$ is the fiducial gas radius for an $L_B^*$ galaxy,
and $\beta$ is a constant power law slope; both are determined
empirically.  Using maximum-likelihood techniques, CT08 estimate
$R_{gas}^*=91 h^{-1}$ kpc, and $\beta=0.35$ for low $z$ \mgii
absorbers.  CT08 then consider the covering fraction within $R_{gas}$
rather than raw impact parameter, finding $\kappa=80 - 86\%$ inside
and $\kappa\approx 0$ outside of $R_{gas}$.  The increase in $\kappa$
from 60\% (uncorrected) to 86\% (corrected) reflects the fact that
many of their \mgii non-detections pair with galaxies of low
luminosity.  These systems' impact parameters, though less than 100
kpc, still fall outside their (smaller) gas radii.

\begin{figure}
\includegraphics[scale = 0.5, clip=true]{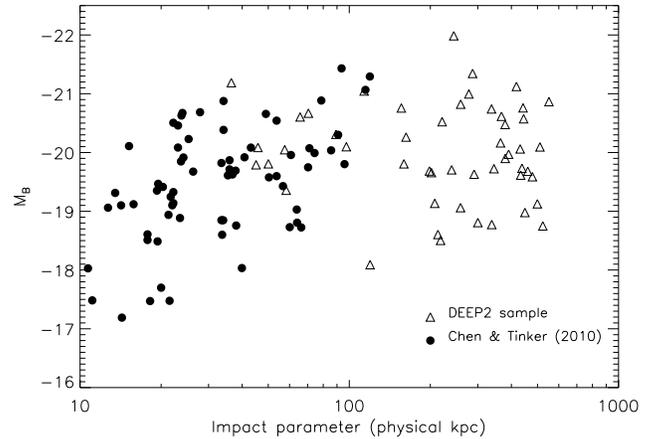}
\caption{Sample comparison between DEEP2 and CT2010, in the impact
  parameter versus absolute magnitude plane.  The CT2010 sample probes
  further down the luminosity function with particular emphasis on
  faint systems with $b \le 20$ kpc that are missed in the high redshift
  sample.}
\label{fig:b_MB}
\end{figure}

At DEEP2 redshifts, the situation is reversed because our galaxies are
drawn from the brighter end of the luminosity function and therefore
reside in larger gas haloes.  This is evident from Figure
\ref{fig:b_MB}, which illustrates the distribution of impact parameter
and galaxy luminosity occupied by the low and high redshift samples.
As expected, the more nearby surveys are weighted towards
low luminosity galaxies at small $b$.

\begin{figure}
%\epsscale{0.8}
%\plotone{f15.eps}
\includegraphics[scale = .5, clip = true, trim = 50 100 0 0]{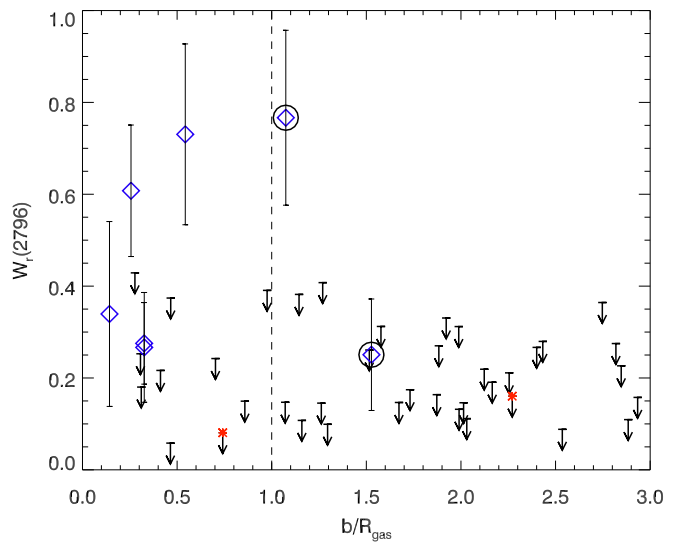}
\caption{\mgii rest equivalent width versus impact parameter in units
  of $R_{gas}$, as defined in Equation 1.  The horizontal axis has
  been truncated at 3.0; non-detections in the DEEP sample extend out
  to $b/R_{gas}$ = 12.  Diamonds (blue) mark detections, while arrows
  represent upper limits.  Detections with black circles are each
  co-located in redshift with another sample galaxy having
  $b/R_{gas}<1$.  Asterisks (red) mark AGN.}
\label{fig:ct_rgas}
\end{figure}

In the $z\sim 1$ sample, we find several non-detections located at $b
> 100$ kpc (i.e. beyond the approximate gas radius for an $L*$
galaxy), but which still fall within $R_{gas}$ because the galaxies
are bright.  This can be seen in Figure \ref{fig:ct_rgas}, where we
plot \mgii $W_r$ against impact parameter expressed in dimensionless
units of $R_{gas}$.  When evaluating $R_{gas}$ for each system, we
correct absolute magnitudes to the rest frame using {\tt kcorrect} and
assume $M_B^*=-19.8+5\log(h)$ for consistency with the lower redshift analysis.
With this adjustment, the covering fraction at $z=0.9$ decreases to
$\kappa=38\%$ (5 detections and 8 non-detections within the new
$R_{gas}$).

\begin{figure}
%\plotone{f16.eps}
\includegraphics[scale = 0.5]{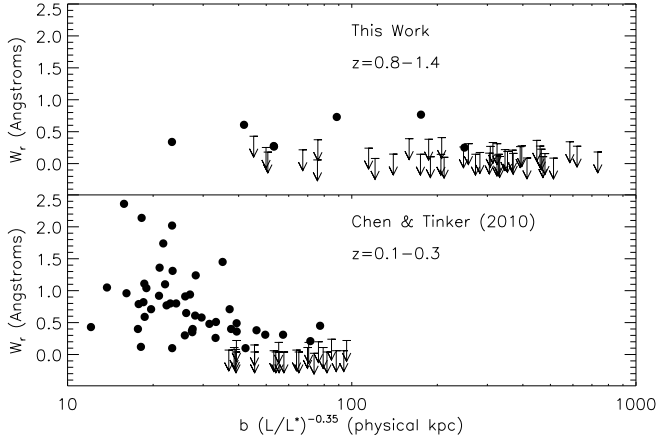}
\caption{Equivalent width versus (luminosity-rescaled) impact
  parameter for the present sample, compared with that of
  \citet{chen2010lowz}, illustrating how the high redshift sample
  skews toward larger impact parameters.  Our innermost point has
  $W_r$ near the low end of CT2010's envelope, but the samples are
  otherwise fairly similar.  The strongest absorbers in the low
  redshift sample all arise in systems with rescaled impact parameters
  smaller than the bulk of the DEEP2 sample.  (Note: our point near $b$ = 50 kpc represents two detections with similar $b$ and $W_r$, not separable on the scale of this plot).}
\label{b_scaled}
\end{figure}

The lower covering fraction we observe in the high-$z$ sample results
at least in part from the fact that these galaxies are located on
average at larger impact parameter, where the \mgii $W_r$ and covering
factor {\em should} be lower.  Figure \ref{b_scaled} illustrates this
graphically: the high covering fractions of the CT08 and CT2010
samples are driven by points at rescaled impact parameter $<20$ kpc,
where the detection probability is essentially unity.  Our sample only
has one system in this range, and although its $W_r$ is in the bottom
10\% of the $W_r$ distribution for $b_{scaled}<30$ kpc in CT2010, it
still represents a positive detection.

The rest of the high redshift sample is at $b(L/L*)^{-0.35}>40$ kpc,
where the \mgii yield is also low in the low redshift sample.  It is
clear that the probability of detecting a \mgii absorber varies
strongly even inside $R_{gas}$.  This implies that cumulative
estimates of $\kappa$ within a fixed radius depend strongly on the
radial selection function of a given sample.  Indeed comparisons of
$\kappa$ from one study to another may not be meaningful if not
corrected for this effect.

\begin{figure}
%\plotone{f17.eps}
\includegraphics[scale = 0.5, clip=true]{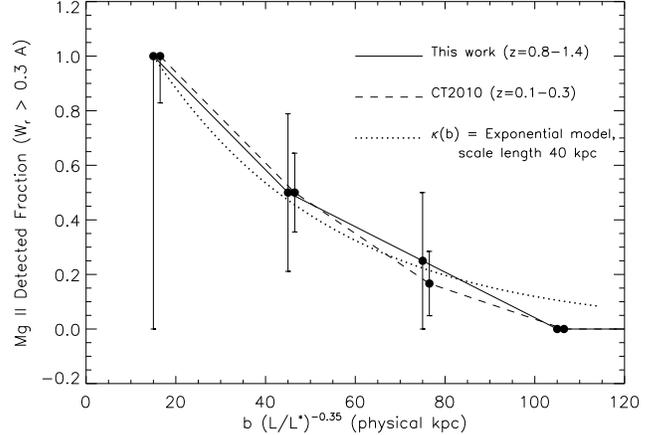}
\caption{Estimates of the differential covering fraction in radial
  bins, for high and low redshift samples.  This quantity expresses
  the probability of detecting \mgii coincident in redshift with a
  galaxy as a function of the galaxy-QSO impact parameter.  It is a
  steeply declining function of radius, with a near unity probability
  at 10-20 kpc dropping to nearly zero at 100 kpc.  This figure
  emphasizes the importance of radial selection functions in
  calculating meaningful cumulative covering fractions within an
  arbitrary radius.  We find strikingly similar scalings at $z\sim
  0.1$ and $z\sim 0.7$, albeit with very large errors at high
  redshift.  Note that in order to avoid double-counting detections associated with pairs of galaxies, we place these detections in the bin corresponding to the $b$ of the closer galaxy.}
\label{kappa_binned}
\end{figure}

Figure \ref{kappa_binned} shows the covering fraction for systems
grouped into radial bins of 30 kpc.  The bin size and shot noise in
each bin is large on account of small sample size, particularly at
high redshift.  But in this view, we see that {\em at fixed galaxy
  luminosity and impact parameter}, the covering fraction of \mgii
absorbers appears to be very similar at $z\sim 0$ and $z\sim 1$,
albeit with substantial uncertainties.

The \mgii covering fraction may be expressed very roughy in terms of
galaxy impact parameter and luminosity:

\begin{equation}
\kappa(b,L) = \exp\left[{-b(L/L^*)^{-0.35} - 15}\over{40}\right]
\end{equation}

This relation is completely empirical, and is not a detailed fit
(given the large uncertainties due to sample size).  However it does
provide some basis for assessing the likelihood that a random galaxy
near a QSO sightline will produce \mgii absorption.  It also allows
for a comparison of samples drawn with different selection functions
in luminosity and impact parameter.

\subsubsection{Trends with Star Formation Rate}

\begin{figure}
%\epsscale{0.8}
%\plotone{f18.eps}
\includegraphics[scale=0.48, clip = true, trim = 30 100 0 0]{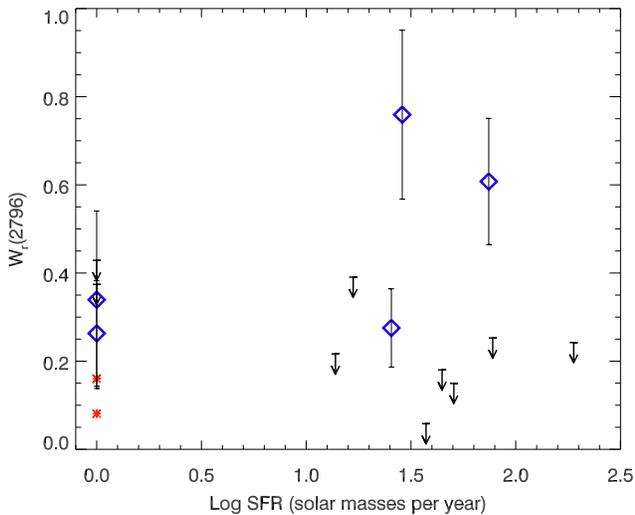}
\caption{\mgii EW vs. log(SFR).  The open diamonds (blue in the color
  figure) represent detected \mgii absorption.  The arrows (black)
  represent upper limits for \mgii non-detections.  Asterisks (red)
  mark AGN candidates.  Only galaxies with impact parameter $b < 100$
  kpc are shown.  See the electronic edition for a color version of
  this figure.}
\label{sfr_vs_mg}
\end{figure}

Figure \ref{sfr_vs_mg} shows a plot of \mgii equivalent width against
star formation rate as determined by the methods discussed in section
\ref{sfr}.  Galaxies with impact parameters $b>200$ kpc are not
included for clarity.  Although our two strongest absorbers are
associated with star forminng systems (30-100 $M_\odot$/year), there
are several other galaxies in the sample with similar SFR but no
detected \mgii halo absorption.  The systems in our sample lacking
emission lines (shown at zero SFR) do harbor comparatively weak \mgii
absorbers.  In general, any trends in this space are too weak to be
revealed by our small sample size.

It may be that passive galaxies are under-represented in the DEEP2
parent sample---especially at high redshift---owing to the photometric
preselection and relative redshift success rates.  However, as noted
by \citet{yan_poststarburst}, there are substantial populations of the
so-called K+A galaxies in $z\sim 1$ samples, indeed we see several
such post-starburst systems in the DEEP archive, many with
non-negligible \oii emission.  None are present in our modest \mgii
sample, but these galaxies may be a very interesting sample to compare
with lower redshifts where they are relatively more rare.

\subsection{Trends with Galaxy Type}

Within our small sample, we do not detect evidence that \mgii
absorption is more commonly associated with either young star forming
galaxies or late type stellar populations.

Three of the galaxies with \mgii halo absorption (DEEP31050080,
31052516, and 31050146) display 4000 \AA\ breaks and little to no \oii
or \hbeta emission (Figure \ref{detected_spectra}).  The four
remaining galaxies range from moderate to strong star formers, based
in part on their \oii line emission and weak 4000\AA\ breaks.  Two
pairs of these galaxies are likely grouped together based on redshift
and spatial distribution: the early types 31050080 and 31050146; and
the emission line galaxies 43056587 and 43056310.  The spectra of
seven galaxies with $b\lesssim 100$ kpc but no \mgii detection are
shown in Figure \ref{nondetected_spectra}.  These 7 galaxies also cover a range of spectral
types: five display \oii emission; DEEP 31020310 is low-redshift and
does not cover \ion{O}{2}, but does show moderate \hbeta and
\ion{O}{3}; and DEEP 32024758 shows a 4000 \AA\ break.

While we stress that our galaxy sample is unbiased {\em with respect
  to DEEP2}, it is important to consider the effects of any bias
inherent in the parent sample.  DEEP's photometric pre-selection and
apparent magnitude limit, coupled with its improved redshift success
rate for emission line galaxies, tends to select star forming systems,
particularly at the high redshift end of the sample \citep{gerke2007}.

\begin{figure}
\epsscale{1.3}
\plotone{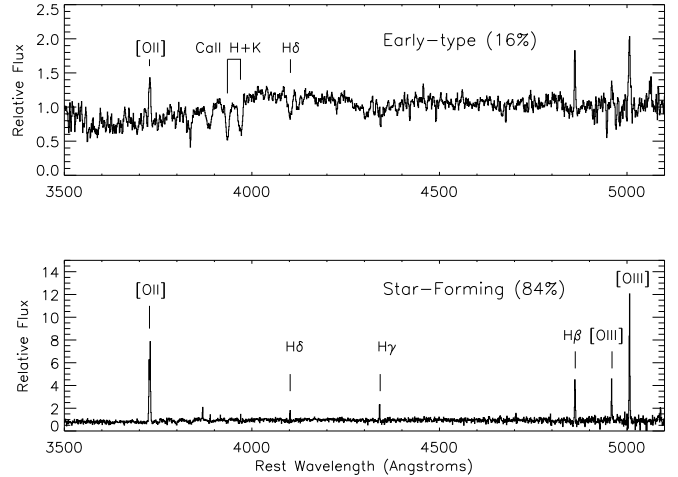}
\caption{Stacked spectra of DEEP2 galaxies classified as Early-type
  (top) and star-forming (bottom).}
\label{stacks}
\end{figure}

To compare the general DEEP2 sample with our \mgii host sample, and
also with the general galaxy population, we drew 1000 spectra from the
DEEP archive to examine the relative frequency of early- versus
late-type galaxies.  Spectra with redshifts outside our selection
range or in the EGS were not examined.  We visually classified each
spectrum as either ``emission-line dominated,'' or ``Early-type.''
The emission systems displayed strong \oii, \oiii, or H$\beta$ lines
but weak or blue continuum, while the continuum systems showed weaker
\oii but stronger 4000 \AA ~breaks and Ca H+K absorption.
Uniform-weighted rest-frame stacks of the two samples are shown in
Figure \ref{stacks}.  We found that 84\% of the DEEP2 galaxies fell in
the emission-dominated category, while 16\% were continuum dominated.

In the \mgii-hosting subsample of DEEP2, three of the seven systems
(42\%) are early-type classifications.  This is a modest excess, but
we are still in the regime of small number statistics.  The Poisson
probability of detecting three early-type \mgii hosts (when a randomly
drawn sample has an expectation value of slightly over one), is 8\%.
It appears the \mgii hosts do not differ strongly from the general
DEEP2 population.  While many \mgii are associated with star-forming
galaxies, they certainly do not appear to uniquely be so.

It is less straightforward to quantify how the DEEP2 sample may be
biased with respect to a volume limited sample of galaxies at $z\sim
1$.  \citet{norberg_2df} performed a similar binary classification of
local 2dF galaxies into early- and late-types using principal
component analysis, finding roughly equal numbers of the two at our
$M^*-5\log h = -19.8$.  Recall that the DEEP limiting magnitude varies
between -18.6 and -20.6 across its full redshift interval.  Toward the
bright end of our absolute magnitude range, the 2dF late-type fraction
decreases to $\sim 30\%$, whereas at the faint end it increases to
$70\%$.  However one expects the fraction of blue galaxies to increase
toward higher $z$, particularly in the group environment
\citep{butcheroemler, gerke}.

In short, we find that \mgii absorption is not exclusively associated
with a particular spectral type in the DEEP2 galaxy sample.  Emission
line galaxies and early type objects alike exhibit \mgii absorption in
some cases and not in others.  A determination of the exact fraction
of early-type versus star forming galaxies hosting \mgii requires an
in-depth analysis of sample bias inherent to DEEP2's selection
process; this is beyond the scope of this paper but will be an
important question to address when larger samples of \mgii-galaxy
pairs are available at $z=1$.  It appears that ongoing star formation
is not a prerequisite for populating $z\sim 1$ circumgalactic haloes
with cool (i.e. $10^4$ K), metal enriched gas; this applies at least
for the intermediate values of $0.3<W_r<1$ \AA ~typical of
galaxy-selected \mgii systems (see Section 4.1).

\subsection{Absorbers Without Known Galaxies}
Five \mgii systems were found in our QSO spectra with redshifts within
DEEP's selection range ($0.7 < z < 1.4$) but no corresponding match in
the DEEP galaxy catalog.  We were interested to know if every \mgii
absorber could be plausibly associated with a luminous galaxy not
captured by DEEP2, or whether some could instead trace galaxies of
very low $L$ or intergalactic gas.  We therefore searched the DEEP
photometric catalogue \citep{coil_deep2} for objects that were
detected in imaging but not included in the redshift catalogue because
of slit collisions or spectroscopic magnitude limits.

First, we applied the same criteria used by the DEEP survey to select
galaxies in the $0.7<z<1.4$ range: color cuts of $B - R < 2.35(R - I)
- 0.25, R - I > 1.15, $ or $ B - R < 0.5$; magnitude cuts of 18.5 $\le
m_B \le$ 24.1; a surface brightness cut of $SB = R_{AB} +
2.5$log$_{10}[\pi(3r_g)^2] \le 26.5$; and a probability of being a
galaxy $p_{gal} > 0.2$.  We assumed objects that meet these conditions
but have no catalog entry were omitted because of slit conflicts.

We eliminated all objects with QSO impact parameter $b \le 1''$, as
these are likely the QSO itself \footnote{One such object, 32024971,
  at $b = 0.365''$, also had redshift and spectral information in the
  DEEP archive and was confirmed as the QSO SDSS233050+0015}.  We then
assumed that every remaining candidate was located at the redshift of
the absorption doublet and calculated its physical impact parameter.
Two such candidates fell within a 150 kpc radius.  They are listed in
Table 4.  Only one of these seems to be a plausible pair at $b<100$ kpc.

We then removed the apparent magnitude cut to search for candidates
that passed the DEEP color cuts but were deemed too faint for
followup, and repeated the calculation.  The eleven candidates found
this way are listed in Table 5.

Of the four QSO fields containing ``bare'' \mgii detections, two had
candidates not observed because of slitmask incompleteness.  All four
fields also had at least one candidate which passed color cuts but was
too faint for followup.  It is therefore possible that every \mgii
system seen in our MagE spectra traces an observable galaxy; i.e. that
there are no ``bare'' \mgii systems.  

The one possible exception to this statement is the $z=0.615$ absorber
in SDSS2230+0015 which shows a weak \mgii system ($W_r=0.148$ \AA) but
no confirmed galaxies.  A pair of faint photometric candidate galaxies
separated by $3.8\arcsec$ is located at an impact parameter of
$21\arcsec$ from the QSO.  If these are positioned at the redshift of
the absorber, they would pass $b=143$ kpc from the line of sight, with
$18.2h^{-1}$ kpc projected separation and have $k$-corrected absolute
magnitudes $M_B=-16.8$ and $-17.8$ ($L_B=0.04-0.16 L^*$).  A third
galaxy, also at $b=21\arcsec$ but far from the other two, would have
$M_B=-16.4$ ($L_B=0.04L^*$).  No additional galaxies with
$L_B>0.1L^*$ were found within 200 kpc.  It appears therefore that
this absorber lacks either a faint host galaxy at low impact parameter
(i.e. fainter than $\sim 0.05L^*$), or a bright galaxy which could
potentially harbor \mgii at large radius.  If {\em both} of the
angular pair members are confirmed to lie at the absorber redshift it
could provide evidence for gravitational interaction as a mechanism
for transporting \mgii to large radii, but without spectroscopic
redshifts this interpretation remains speculative.

All other fields all have candidates within $100$ kpc, and in some
cases several candidates.  Followup observations would be required to
establish the redshifts of these objects and either confirm or rule
out their association with the \mgii absorbers.

In principle one can make qualitative statements about the types of
galaxies these candidate galaxies might be based upon DEEP2's three
color photometry and knowledge of the absorber redshift.  However,
since each of our ``bare'' \mgii systems has multiple photometric
candidates for the host, one cannot make strong statements about the
relative contributions of early versus late type galaxies to these
unidentified systems.  

Nevertheless, we did examine the color properties of each candidate by
calculating {\tt kcorrect} templates for each system, placed at the
redshift of its corresponding ``bare'' \mgii absorber.  Qualitatively,
for galaxies near DEEP2's central redshift, the $(R-I)$ color measures the
strength of the $4000$ \AA ~break; galaxies with $(R-I)\gtrsim 0.6$
are more likely to have strong breaks and hence be early type systems,
while lower values would select later types.  The strength of the
$(R-I)$ index decreases at the upper end of DEEP's redshift range as
the $4000$ \AA ~break moves throguh the $I$ band filter.  We find two
systems whose $(R-I)$ colors (and {\tt kcorrect} templates) favor old
stellar populations, and six which favor younger populations, not
unlike the main sample.  If one lines up candidates at their putative
redshift (as determined by \mgii) and assumes that candidates map out
a similar region of $(R-I)$ color as the deep sepctroscopic sample,
then certain candidates become more favorable.  When this assumption
is applied, then two ``bare'' \mgii systems are more likely to be
hosted by late-type galaxies ($z=0.684$ in SDSS232634, and $z=1.477$
in SDSS232707), one is more likely an early-type galaxy ($z=0.952$ in
SDSS232707, and two are inconclusive.  At the very least the
photometric candidates do not appear to map out a different portion of
parameter space.

\section{Discussion}
Within our small sample, the properties of galaxies giving rise to
\mgii absorption are heterogeneous.  The likelihood of detecting \mgii
absorption appears to scale only with distance and luminosity, and not
with star formation or stellar population.  We do not see strong
evidence that $W_r$ increases at low $b$, although our coverage is
poor at $b\lesssim 30$ kpc where this effect is most pronounced at low
redshift.  After correcting for galaxy luminosity, the scaling of
covering fraction with impact parameter appears to be very similar
at $z\sim 1$ and $z\sim 0$.

While our galaxy population is mixed, none produce absorbers stronger
than $\sim0.8$ \AA .  Even including absorbers not associated with a
DEEP galaxy, our strongest detected absorber has a $W_r$ of only
1.0\AA\ , while our weakest is $\sim0.15$\AA.  Previous studies that
have identified \mgii absorption with star forming galaxies have also
generally selected for high-strength \mgii lines; none of our systems
would have been included in these samples.  For example,
\citet{bouche2007ha} found evidence of \ha emission correlating to
\mgii absorption, but limited the sample to absorbers of $W_r > 2$
\AA.  \citet{menardo2}, which correlated \mgii absorption with \oii
luminosity density, used a lower bound on absorber strength of 0.7\AA.
\citet{zibetti2007stacking} demonstrated a systematic excess in blue
luminosity around quasar sightlines with \mgii absorbers, but again
used a lower bound on absorber strength of 0.8\AA.

By contrast, studies dominated by weaker absorbers (i.e.  $W_r \le
1.0$\AA) do not seem to require this connection between \mgii and star
formation in all systems \citep[e.g.][]{chen2010lowz}.  Our \mgii
sample compares more closely with this latter group, and our findings
are similar.  These differing results suggest a divide in the nature
of high-strength vs. low-strength absorbers.  It may be that
high-strength absorbers are direct byproducts of star formation while
low-strength absorbers are not.  Several authors have previously
considered the possibility of multiple \mgii populations:
\citet{churchill_zoo} segregated these populations by their associated
high-ionization absorption, while \citet{nestor2005} showed tentative
statistical evidence of a discrepancy in redshift evolution between
strong and weak systems.

\subsection{A Weak Evolution in Covering Fraction: Interpretations}

Samples of strong \mgii absorbers ($W_r > 1$ \AA)
show some evidence of association with star formation, while lower
$W_r$ samples like ours do not show such a strong trend.  It may be
that the strong absorbers are a direct consequence of outflows during
the actual starburst/outflow phase, while weaker absorbers represent
the dispersed remnants of this phase as they settle into the galactic
halo.  The dispersal could also be aided by tidal disruption or
dynamical processes during the halo merging process.

In this picture, the weaker absorbers found in galaxy-selected surveys
such as the DEEP2 sample, CT08, and CT2010, represent relics from
either earlier episodes of star formation, halo assembly, or both
processes together.  This would explain the existence of these systems
around both early type and late-type galaxies.  The strong systems
found in absorber-selected surveys such as \citet{bouche2007ha},
\citet{menardo2} and \citet{zibetti2007stacking}, on the other hand, would tend to select
more active star forming galaxies.

This interpretation would also explain the redshift evolution in
classical, purely statistical \mgii absorber surveys \citep[e.g.][]{nestor2005}, which
show a rapid increase in ``strong'' absorbers toward higher redshift
with a gradual {\em decrease} in $W_r=0.3-1.0$ \AA ~absorbers over the
same interval.  Strong absorbers would track the instantaneous star
formation rate and would naturally be more common during periods of
intense star formation.  The weak absorbers, by contrast, would gradually accumulate
in number with time (toward lower redshift), as they track the
integrated star formation rate and/or halo mass assembly.  This model
also explains the order of magnitude prevalence of weak absorbers over
strong ones at all redshifts.

The association of the $W_r<1$ \AA ~systems with relic halo gas may
account for the relatively weak change in covering properties observed
between our $z\sim 1$ sample and similarly constructed $z\sim 0.2$
samples.  While CT08 treat \mgii absorption as a generic feature of
gaseous halo assembly, they also examine the anticorrelation of halo
mass and \mgii $W_r$ \citep{bouche2006anticorrelation, gauthier}.
Models favoring outflow origins of \mgii explain this anticorrelation
as a byproduct of lower mass galaxies having higher star formation and
outflow rates; CT08 interpret it as a signature of cold mode accretion
\citep{dekel} of halo gas for low mass systems transitioning to hot mode
for higher galaxy masses.

Naively, one would expect that at higher redshifts cold mode accretion
would be more prevalent and therefore the covering fraction of \mgii
would increase, contrary to our findings.  A detailed answer to this
question is probably not possible without comparison to simulations,
which are challenged to resolve cold flows at redshifts one and zero.
Recently \citet{faucher_coldflows} simulated individual galaxies at
high resolution to examine the covering fraction of \ion{H}{1} at
$z=2-4$.  Their calculations showed covering of only a few percent for
Damped Lyman-alpha column densities.  However for the equivalent of
Lyman limit systems---which may be a more fair comparison for \mgii
absorbers---the covering fractions range from 10-15\%.  The
calculations did not extend to lower redshifts appropriate for our
sample, although the trend indicates that the fraction increases
from $z=2$ to $z=4$.  

One would also expect the overall observed \mgii covering fraction to
increase if all \mgii systems are related to star formation.  This is
particularly true of the strong \mgii systems in the picture we have
described above.  We have not found such evidence, indicating that our
selection method may be less disposed to select \mgii arising from
starburst outflows, relative to absorption-line selected galaxy
samples.

\citet{chenandtinker2009} examine the redshift evolution of the
covering fraction using halo occupation models coupled to analytic
models of cold accretion.  They find a strong redshift dependence of
$\kappa\propto(1+z)^{2.75}$ for the {\em total} \mgii covering from
haloes of all galaxy masses.  The increase in $\kappa$ with $z$ is
much stronger for haloes below $\sim 10^{11.5}M_\odot$, while more
massive haloes do not evolve strongly.  We are almost certainly
sampling the high end of this mass distribution at high redshift, but
we do not see a strong rise in \mgii number counts from undetected low
mass galaxies (see Section 3.6).  However, if followup spectroscopy
demonstrates that the \mgii systems without known galaxies are not
paired with any of our photometric candidates---especially the
brighter ones missed from slit collisions---then any host galaxy would
have to have very low $L$, lending support to this picture.

\subsubsection{Photoionization of \mgii Systems}

Broadly, both the cold accretion and starburst outflow would predict
an increase in \mgii covering at high redshift which we do not
observe.  One effect which could counteract this tendency, and has not
been discussed extensively in the context of evolution, is the
extragalactic ionizing background.  The intensity of the UV radiation
field rises with increasing redshift between $z=0$ and $z=1$, and can
eliminate \mgii by ionizing it further to \ion{Mg}{3}.

For the galaxies in our sample, we calculated the ratio of local to
extragalactic ionizing flux at their individual impact parameter,
using each galaxy's measured absolute magnitude and best fit {\tt
  kcorrect} model spectral template along with a \citet{giguereuvbg}
ionizing background spectrum.  In all cases the extragalactic flux (at
the \mgii ionization edge) dominates the illuminating radiation field
of the galaxy, by factors ranging from 10 to 10,000, with a median
ratio of $\sim 1000$.

The degree to which this boost in the UV background ionizes away \mgii
depends on the density of the systems as well as their hydrogen column
density.  We ran simple CLOUDY models to examine this effect over a
range of number density, and found that for densities $n_H > 0.05$
cm$^{-3}$ the \mgii equivalent width is mostly unchanged; but for
lower values ($n_H\sim 0.001-0.01$), the difference in background flux
between $z\sim 0$ and $z\sim 1$ could reduce $W_r$ by a factor of 2-3.

The effect of the UV background on $N(z)$ for \mgii could be reduced
if the absorbers are optically thick in \ion{H}{1}, since the \mgii
ionization potential is at only 1.1 Ryd and there is some evidence
that \mgii absorbers like the ones seen here relate closely to Lyman
limit systems or the outskirts of Damped Lyman alpha systems.  For
typical column densities seen in LLS or DLA absorbers ($N_{\rm HI}\sim
10^{18}-10^{20}$) the densities tested above imply an absorber
thickness of order only 100 pc to one kpc.  These would seem quite
small, but are in fact similar in scale to the sizes of \mgii systems
inferred from ionization modeling of individual systems
\citep{charltonetal03} and observations of multiply-imaged
gravitational lenses intersecting \mgii haloes \citep{rauch}.

In short, changes in the ionization of \mgii could help explain why
the covering fraction at $z\sim 1$ is similar to that at $z\sim 0.2$,
despite general expectations that it should rise in cold accretion or
outflow models.  Ionization and/or photoheating may play a larger
role at early times resulting in a small net evolution in $\kappa$ and
$N(z)$ for systems with $0.3<W_r<1.0$ \AA.  Surveys for \mgii at
$z>2-3$ (past the peak in the UV background) should help to
discriminate between the evolutionary effects of the UV background and
gas physics. 

\subsection{Comparison to $dN/dz$ Estimates for \mgii Systems}

We briefly consider the weak evolution in \mgii-galaxy properties in
the context of traditional, blind \mgii absorption studies.  These
large surveys repeatedly find that $dN/dz$ for \mgii {\em increases}
with $z$ \citep{nestor2005, prochtersdss2006, steidelandsargent}
slightly faster than a population of sources with constant comoving
density.  This requires that either the covering fraction $\kappa$
increase on a per-galaxy basis, or the number of candidate galaxies
increase at earlier times.

\citet{nestor2005} divide their SDSS \mgii sample into bins of $W_r$
and examine each bin's evolution separately with respect to the
non-evolving case.  They find that the majority of the SDSS absorber
sample {\em is} broadly consistent with no-evolution, and that overall
evolution is driven primarily by movement at the strong ($W_r>3.0$
\AA) and weak ($W_r<0.6$ \AA) ends of the $W_r$ distribution.  At the
strong end, which is not represented in our sample, $dN/dz$ increases
toward higher redshift much faster than a non-evolving population;
this has been discussed here as well as by Nestor and others in
connection with star formation and galactic outflows.

At the low $W_r$ end more appropriate for our sample, Nestor et al.
find evidence at the $2\sigma$ level for a {\em decline} in
$dN/dz$---by a factor of $\sim 2$ for $0.3<W_r<0.6$ and $z=0.5$ to
$z=1$.  These results have marginal statistical significance, but they
do demonstrate that in the range where our survey is sensitive,
evolutionary effects are substantially smaller in weak vs. strong populations in both blind
absorption and galaxy-selected surveys.

\subsection{Very recent work on \mgii at $z\sim 1$}

During review of this paper, \citet{lundgren} presented a
complementary analysis of strong \mgii absorbers around DEEP2 galaxies
as measured directly from SDSS survey quasar spectra.  With a limiting
equivalent width of $W_r>0.6$ \AA, this survey is somewhat less
sensitive to \mgii absorption and focuses primarily on large scale
clustering analysis.  Their group reports a 50\% covering fraction
within 60$h^{-1}$ kpc (one of two possible systems) and lower covering
at larger impact parameters.  They detect a large galaxy absorber pair
in the spectrum of SDSS0228+0027 (DEEP41013534) with impact parameter
42 $h^{-1}$ kpc not included in our sample because of priors imposed
on observations of background QSOs (the background QSO is at $z=2.4$).
This galaxy is a luminous, clearly early-type system.  It further
underlines that a non-negligible fraction of \mgii absorbers may be
associated with galaxies having low instantaneous star formation
rates.  Interstingly, this system, like the $z=0.790$ absorber we
report in SDSS2325+0019, may reside in a galaxy group.

In a second recent paper, \citet{bordoloi} examined stacked spectra of
$\sim 100$ foreground-background galaxy pairs in the lower half of our
redshift range.  While not appropriate for covering fraction
measuremenets, this analysis does produce a mean \mgii equivalent
width measurement (for the combined $2796$\AA ~and 2803 \AA
~components), which traces the combined effects of absorber strength
and covering fraction.  The stacked spectrum for all ``red'' galaxies,
which are selected to resemble the early-type systems reported here,
yields low $W_r$ for all impact parameters probed, a factor of
2-3$\times$ below our median, while the ``blue'' population shows very
strong absorption that rises steeply with decreasing radius and
increasing stellar mass.

The trends in Bordoloi et al's blue sample are crudely consistent with
our findings and those of \citet{lundgren} as well as
\citet{chen2010lowz}.  Their declining stacked $W_r$ at high $b$ may
be interpreted as a drop in covering fraction toward higher impact
parameters, coupled with a somewhat smaller effect from a decreasing
average $W_r$ of each absorber.  Comparison of our Figures 16 and 17
roughly reproduce the scalings in Bordoloi's Figure 4.

The apparent weak absorption from ``red'' galaxies at all impact
parameters is less consistent with the aforementioned galaxy-selected
samples.  The cause of this discrepancy is not immediately clear; a
full accounting will require either larger samples of individual
early-type galaxy/absorber systems to measure a detailed covering
fraction, or else a separate stack of higher-resolution galaxy spectra
to test whether this result based on background galaxies, rather than
QSOs, continues to hold.

\section{Conclusions}

We have presented a new sample of 10 QSO spectra selected randomly
for their proximity to DEEP2 survey galaxies to examine the frequency
of \mgii absorption in extended gas haloes at $z\sim 1$.  These
objects, which each have at least one DEEP2 galaxy at $b<100$ kpc (and
many more at larger $b$), were observed at medium resolution with the
MagE spectrograph.  The galaxies have $\bar{z}=0.87$ and were not
chosen for any particular intrinsic properties save for those
inherent to the DEEP survey selection criteria.

Our findings may be summarized as follows:
\begin{enumerate}
\item{\mgii is found in the haloes of early-type and late-type
  galaxies alike at $z\sim 1$.  We see no suggestion of a bias toward
  any particular stellar population or instantaneous star formation
  rate.  Neither of the two possible AGN in the sample exhibits
  associated \mgii absorption.}
\item{We find 5 unique galaxy-\mgii pairs, two of which are grouped
  with a second galaxy at similar redshift.  Five galaxies with DEEP2
  spectra at $b<100$ kpc do not have an associated \mgii absorber.}
\item{We find five additional \mgii systems in our spectra with no
  associated galaxy in the DEEP2 catalog.  However in all but perhaps
  one case, a plausible candidate was recovered from the DEEP
  photometric catalog.  These candidates have small impact parameters,
  but were not observed because of slit mask constraints or a narrow
  miss of the DEEP2 apparent magnitude cutoff.}
\item{After applying a simple scaling for galaxy luminosity
  (i.e. accounting for the idea that more luminous galaxies have more
  extended haloes), the derived dependence of covering fraction with
  galaxy impact parameter at $z\sim 1$ becomes strikingly similar to
  what is observed in similar samples at $z\sim 0.2$.  This highlights
  the importance of accounting for radial sampling functions when
  calculating the covering fraction of \mgii or other ions within a
  fixed impact parameter.}
\end{enumerate}

The \mgii systems in our sample fall between the regime of the
``weak'' systems with $W_r<0.3$\AA ~\citep{churchill_weakmgii}, and the
``strong'' systems with $W_r>1$ \AA, the latter of which several
groups have suggested may trace galactic outflows. We do not see a
strong link between star formation and \mgii incidence, but rather
find systems around a wide range of spectral types.  If our
intermediate-strength \mgii systems relate to outflows, they may be
relic metal-rich material accumulated by either past star formation
episodes or dynamical processes occurring during galaxy assembly.

Both the outflow and cold accretion hypotheses would predict an
increase in \mgii covering fraction with redshift, which we do not
observe.  Halo occupation models predict weak evolution for haloes
with mass $M>10^{12}M_\odot$, representative of our sample.  However
they also predict a sharp rise in \mgii absorption from low mass
haloes at higher $z$ which we do not conclusively see.  We speculate
that the changing extragalactic UV background may play a role in
suppressing \mgii absorption at high $z$, provided 1.1 Ryd photons can
penetrate the structures where the absorption occurs.

Observations of \mgii absorption at $z>2$ may hold some insights into
the relationship between halo assembly, star formation, feedback, and
ionization evolution.  Beyond this redshift, the ionizing background
once again declines, which should favor \mgii in the ionization
balance.  It remains to be seen how this trades off against the
decline in abundances at early times.  Moreover it is not clear that
galaxy scaling relations for $L^*$ systems derived locally would be
well justified at such early times.  We are exploring this further
using IR observations which will extend \mgii number density surveys
to $z>4$.

\bigskip

{\it Facilities:} \facility{Magellan:Clay (MagE)}

\acknowledgements

We thank Hsiao-Wen Chen for helpful conversations and sharing her data
in electronic form, and for conversations with DEEP2 team members,
particularly David Rosario.  Thanks to Kathy Cooksey for helpful
suggestions on an early draft.  We thank the staff of Las Campanas
Observatory and the Magellan Telescopes for their assistance obtaining
the data used here.  We also gratefully acknowledge the support of the
Sloan Foundation.  RAS gratefully acknowledges generous lumbar support
from the Adam J. Burgasser Chair in Astrophysics.  This work was
partly supported by the NSF under grant AST-0908920.

\bigskip

Funding for the DEEP2 survey has been provided by NSF grants
AST95-09298, AST-0071048, AST-0071198, AST-0507428, and AST-0507483 as
well as NASA LTSA grant NNG04GC89G.

Some of the data presented herein were obtained at the W. M. Keck
Observatory, which is operated as a scientific partnership among the
California Institute of Technology, the University of California and
the National Aeronautics and Space Administration. The Observatory was
made possible by the generous financial support of the W. M. Keck
Foundation. The DEEP2 team and Keck Observatory acknowledge the very
significant cultural role and reverence that the summit of Mauna Kea
has always had within the indigenous Hawaiian community and appreciate
the opportunity to conduct observations from this mountain.

Funding for the SDSS and SDSS-II has been provided by the Alfred
P. Sloan Foundation, the Participating Institutions, the National
Science Foundation, the U.S. Department of Energy, the National
Aeronautics and Space Administration, the Japanese Monbukagakusho, the
Max Planck Society, and the Higher Education Funding Council for
England. The SDSS Web Site is http://www.sdss.org/.

The SDSS is managed by the Astrophysical Research Consortium for the
Participating Institutions. The Participating Institutions are the
American Museum of Natural History, Astrophysical Institute Potsdam,
University of Basel, University of Cambridge, Case Western Reserve
University, University of Chicago, Drexel University, Fermilab, the
Institute for Advanced Study, the Japan Participation Group, Johns
Hopkins University, the Joint Institute for Nuclear Astrophysics, the
Kavli Institute for Particle Astrophysics and Cosmology, the Korean
Scientist Group, the Chinese Academy of Sciences (LAMOST), Los Alamos
National Laboratory, the Max-Planck-Institute for Astronomy (MPIA),
the Max-Planck-Institute for Astrophysics (MPA), New Mexico State
University, Ohio State University, University of Pittsburgh,
University of Portsmouth, Princeton University, the United States
Naval Observatory, and the University of Washington.

\begin{deluxetable*}{cccccccccc}
\tablecolumns{10}
\tablewidth{0pt}
\tabletypesize{\footnotesize}
\tablecaption{\mgii Absorption Line Systems and Matched DEEP2 Galaxies}
\tablehead{
\colhead{Quasar Sightline} &
\colhead{$z_{QSO}$\tablenotemark{1}} &
\colhead{DEEP2 Match\tablenotemark{2}} &
\colhead{$z({\rm MgII})$} &
\colhead{$b$ \tablenotemark{3}} &
\colhead{$M_B$\tablenotemark{4}} &
\colhead{$SFR$\tablenotemark{5}} &
\colhead{$W_r$ \tablenotemark{6}} &
\colhead{$\sigma(W_r)$} &
\colhead{$\Delta v$\tablenotemark{6}}}
\startdata
SDSS0231+0052 & 1.61 & \nodata  & 0.820 & \nodata & \nodata & \nodata &0.298 & 0.053 & \nodata \\
              &      & 43056587 & 1.009 & 100.02  & -20.13  & 26      & 0.759 & 0.064 & 11      \\
              &      & 43056310 & 1.009 & 226.90  & -20.56  & 112     & 0.759 & 0.064 & 164     \\
\hline
SDSS2325+0019 & 1.21 & 31050080 & 0.790 & 72.09   &  -20.70 & 0       & 0.263 & 0.040 & -150    \\
              &      & 31050146 & 0.790 & 345.24  & -20.77  & 0       & 0.263 & 0.040 & 387     \\
\hline
SDSS2326+0021 & 1.25 & \nodata  & 0.684 & \nodata & \nodata & \nodata & 1.041 & 0.061 & \nodata \\
%              &      & \nodata  & 0.339 & \nodata & \nodata & &0.296 & 0.138 & \nodata \\
              &      & 31052516 & 0.735 & 37.25   & -21.21  & 0       &0.339 & 0.067 & -266    \\
\hline
SDSS2327+0003 & 1.75 & \nodata  & 0.952 & \nodata & \nodata & \nodata &0.221 & 0.037 & \nodata \\
              &      & \nodata  & 1.477 & \nodata & \nodata & \nodata &0.419 & 0.038 & \nodata \\
\hline
SDSS2330+0008 & 1.00 & 32014809 & 0.870 & 58.95   & -20.08  & 29      &0.275 & 0.030 & 420     \\
%              &      & \nodata  & 0.407 & \nodata & \nodata & &0.105 & 0.023 & \nodata \\
%              &      & \nodata  & 0.300 & \nodata & \nodata & &0.295 & 0.030 & \nodata \\
\hline
SDSS2330+0015 & 1.94 & \nodata  & 0.615 & \nodata & \nodata & \nodata &0.148 & 0.026 & \nodata \\
%              &      & \nodata  & 0.309 & \nodata & \nodata & &0.277 & 0.031 & \nodata \\
              &      & 32024543 & 1.187 & 47.06   & -20.13  & 24      &0.608 & 0.048 & -95     \\
%              &      & \nodata  & 1.599 & \nodata & \nodata & &0.188 & 0.048 & \nodata \\
%              &      & \nodata  & 1.749 & \nodata & \nodata & &0.574 & 0.032 & \nodata \\
\enddata
\tablenotetext{1}{QSO emission redshift}
\tablenotetext{2}{DEEP2 redshift catalog number for galaxies coincident with $z({\rm MgII})$.  Absorbers with no matched galaxies indicated with an ellipsis.}
\tablenotetext{3}{QSO-galaxy impact parameter, in $h_{71}^{-1}$ proper kpc}
\tablenotetext{4}{Rest-frame $B$ band absolute magnitude}
\tablenotetext{5}{Logarithmic star formation rate, $M_\odot$/year.}
\tablenotetext{6}{Rest-frame \mgii $\lambda2796$ \AA ~equivalent width in \AA.}
\tablenotetext{7}{Absorber-galaxy velocity offset (km/s)}
\end{deluxetable*}

\begin{deluxetable*}{cccccccc}
\tablecolumns{8}
\tablewidth{0pt}
\tablecaption{DEEP Galaxies at $b<250$ kpc with No Associated \mgii}
\tablehead{
\colhead{Quasar Field} &
\colhead{$z_{QSO}$} &
\colhead{DEEP Name} &
\colhead{$z_{gal}$} &
\colhead{$b$} &
\colhead{$M_B$} &
\colhead{SFR} &
\colhead{$W_{lim}$}}
\startdata
SDSS2330+0008 & 1.00 & 32014683 & 0.8049 & 206.57 & -19.69 & 22 & 0.099\\
\hline
SDSS0231+0024 & 1.05 & 42005413 & 0.9108 & 59.72  & -19.39 & 14  & 0.216\\
            &      & 42005658 & 0.8278 & 159.74 & -20.79 & 189 & 0.242\\
            &      & 42005668 & 0.8262 & 166.29 & -20.29 & 50 & 0.149\\
            &      & 42005439 & 0.8371 & 203.12 & -19.72 & 0  & 0.145\\
\hline
SDSS2325+0019 & 1.21 & 31049930 & 1.1573 & 251.06 & -22.03 & 0  & 0.081\\
\hline
SDSS0231+0044 & 1.26 & 43046656 & 1.1661 & 91.68  & -20.35 & 37 & 0.058\\
            &      & 43046720 & 0.7475 & 223.14 & -18.53 & 4 & 0.111\\
\hline
SDSS2326-0003 & 1.27 & 31008442 & 1.2027 & 51.44  & -19.85 & 78 & 0.253\\
\hline
SDSS0231+0052 & 1.61 & 43056496 & 0.8153 & 245.49 & -19.74 & 51 &  0.261\\
\hline
SDSS0226+0043 & 1.66 & 41045894 & 0.8488 & 46.12  & -19.82 & 0 & 0.429\\
            &      & 41046028 & 0.8479 & 212.58 & -19.17 & 0 & 0.312\\
\hline
SDSS2327+0003 & 1.75 & 31020243 & 1.0047 & 67.35  & -20.64 & 45 & 0.180\\
            &      & 31020310 & 0.6620 & 121.97 & -18.12 & 13 & 0.408\\
            &      & 31020364 & 0.8115 & 163.24 & -19.84 & 17 & 0.391\\
            &      & 31019960 & 0.7400 & 218.14 & -18.63 & 0  & 0.331\\
\hline
SDSS2330+0015 & 1.94 & 32024758 & 1.1027 & 116.59 & -21.09 & 0 & 0.374\\
\enddata
\end{deluxetable*}

\begin{deluxetable*}{ccccccccc}
\tablewidth{0pt}
\tablecolumns{9}
\tablecaption{DEEP2 Photometric Candidates in Fields with Unmatched \mgii Systems}
\tablehead{
\colhead{Quasar} &
\colhead{DEEP Object\tablenotemark{1}} &
\colhead{$\theta$\tablenotemark{2}} &
\colhead{$m_B$\tablenotemark{3}} &
\colhead{$(R-I)$} &
\colhead{$M_B$\tablenotemark{4}} &
\colhead{$z_{\rm MgII}$} &
\colhead{$b$\tablenotemark{5}} &
\colhead{$W_{r}$\tablenotemark{6}}}
\startdata
SDSS2326+0021 & 31052514 & 9.7  & 23.59 & 0.32 & -18.45 & 0.684  & 68.901 & 1041 \\
SDSS2330+0015 & 32024973 & 21.1 & 23.88 & 0.57 & -17.83 & 0.615  & 143.509 & 148  \\
\enddata
\tablenotetext{1}{DEEP2 catalog number}
\tablenotetext{2}{Imact parmameter (arcsec)}
\tablenotetext{3}{Measured, observed frame apparent $B$ magnitude}
\tablenotetext{4}{Inferred rest frame, $k$-corrected absolute $B$ magnitude, expressed as $M_B-5\log(h)$, assuming that candidate lies at $z_{\rm MgII}$.}
\tablenotetext{5}{Inferred impact parameter in $h_{71}^{-1}$ kpc, assuming that candidate lies at $z_{\rm MgII}$.}
\tablenotetext{6}{Rest-frame \mgii $\lambda 2796$ \AA ~equivalent width}
\end{deluxetable*}

\begin{deluxetable*}{ccccccccc}
\tablewidth{0pc}
\tablecolumns{9}
\tablecaption{Faint DEEP2/\mgii Photometric Candidates Excluded by Apparent Magntude}
\tablehead{
\colhead{Quasar Field} &
\colhead{DEEP Object\tablenotemark{1}} &
\colhead{$\theta$\tablenotemark{2}} &
\colhead{$m_B$\tablenotemark{3}} &
\colhead{$(R-I)$} &
\colhead{Inferred $M_B$\tablenotemark{4}} &
\colhead{$z_{\rm MgII}$} &
\colhead{$b$\tablenotemark{5}} &
\colhead{$W_{r}$\tablenotemark{6}}}
\startdata
SDSS0231+0052 	& 43056354 & 4.0  & 24.67 & 0.56 & -18.23 & 0.820 & 30.5  & 323 \\
		& 43056414 & 13.3 & 24.21 & 0.00 & -17.71 & 0.820 & 101.5 & 298 \\
SDSS2326+0021	& 31052590 & 18.2 & 24.22 & 0.22 & -17.65 & 0.684 & 129.7 & 1041\\
SDSS2327+0003 	& 31019897 & 2.5  & 25.25 & 0.87 & -18.56 & 0.952 & 20.2  & 665 \\
		& 31020185 & 9.0  & 24.53 & 0.19 & -18.38 & 0.952 & 71.8  & 221 \\
		& 31020316 & 16.7 & 24.94 & 0.05 & -17.60 & 0.952 & 133.5 & 221 \\
		& 31019897 & 2.5  & 25.25 & 0.87 & -20.94 & 1.477 & 21.7  & 419 \\
		& 31020185 & 9.0  & 24.53 & 0.19 & -19.80 & 1.477 & 77.2  & 419 \\
		& 31020316 & 16.7 & 24.94 & 0.05 & -18.91 & 1.477 & 143.4 & 419 \\
SDSS2330+0015	& 32024604 & 21.0 & 24.59 & 0.22 & -16.80 & 0.615 & 143.1 & 148 \\
		& 32024616 & 20.8 & 25.23 & 0.12 & -16.41 & 0.615 & 141.2 & 148 \\
\enddata
\tablenotetext{1}{DEEP2 catalog number}
\tablenotetext{2}{Imact parmameter (arcsec)}
\tablenotetext{3}{Measured, observed frame apparent $B$ magnitude}
\tablenotetext{4}{Inferred rest frame, $k$-corrected absolute $B$ magnitude, expressed as $M_B-5\log(h)$, assuming that candidate lies at $z_{\rm MgII}$.}
\tablenotetext{5}{Inferred impact parameter in $h_{71}^{-1}$ kpc, assuming that candidate lies at $z_{\rm MgII}$.}
\tablenotetext{6}{Rest-frame \mgii $\lambda 2796$ \AA ~equivalent width}
\end{deluxetable*}

\clearpage
\clearpage

\nocite{kcorrect}
\nocite{wmap7}
\nocite{prochtergrbsightlines}
\nocite{jakobssongrbpair}
\nocite{nestor2010strong}
\nocite{chenandtinker2009}
\nocite{chen2010lowz}
\nocite{menardo2}
\nocite{giguereuvbg}
\bibliography{bib_new}{}

\begin{thebibliography}{46}
\expandafter\ifx\csname natexlab\endcsname\relax\def\natexlab#1{#1}\fi

\bibitem[{{Bergeron} \& {Boiss{\'e}}(1991)}]{bergeron_boisse}
{Bergeron}, J., \& {Boiss{\'e}}, P. 1991, \aap, 243, 344

\bibitem[{{Blanton} \& {Roweis}(2007)}]{kcorrect}
{Blanton}, M.~R., \& {Roweis}, S. 2007, \aj, 133, 734

\bibitem[{{Bochanski} {et~al.}(2009){Bochanski}, {Hennawi}, {Simcoe},
  {Prochaska}, {West}, {Burgasser}, {Burles}, {Bernstein}, {Williams}, \&
  {Murphy}}]{mase}
{Bochanski}, J.~J., {et~al.} 2009, \pasp, 121, 1409

\bibitem[{{Bond} {et~al.}(2001){Bond}, {Churchill}, {Charlton}, \&
  {Vogt}}]{bond}
{Bond}, N.~A., {Churchill}, C.~W., {Charlton}, J.~C., \& {Vogt}, S.~S. 2001,
  \apj, 562, 641

\bibitem[{{Bordoloi} {et~al.}(2011){Bordoloi}, {Lilly}, {Knobel}, {Bolzonella},
  {Kampczyk}, {Carollo}, {Iovino}, {Zucca}, {Contini}, {Kneib}, {Le Fevre},
  {Mainieri}, {Renzini}, {Scodeggio}, {Zamorani}, {Balestra}, {Bardelli},
  {Bongiorno}, {Caputi}, {Cucciati}, {de la Torre}, {de Ravel}, {Garilli},
  {Kovac}, {Lamareille}, {Le Borgne}, {Le Brun}, {Maier}, {Mignoli}, {Pello},
  {Peng}, {Perez Montero}, {Presotto}, {Scarlata}, {Silverman}, {Tanaka},
  {Tasca}, {Tresse}, {Vergani}, {Barnes}, {Cappi}, {Cimatti}, {Coppa},
  {Diener}, {Franzetti}, {Koekemoer}, {Lopez-Sanjuan}, {McCracken}, {Moresco},
  {Nair}, {Oesch}, {Pozzetti}, \& {Welikala}}]{bordoloi}
{Bordoloi}, R., {et~al.} 2011, ArXiv e-prints

\bibitem[{{Bouch{\'e}} {et~al.}(2006){Bouch{\'e}}, {Murphy}, {P{\'e}roux},
  {Csabai}, \& {Wild}}]{bouche2006anticorrelation}
{Bouch{\'e}}, N., {Murphy}, M.~T., {P{\'e}roux}, C., {Csabai}, I., \& {Wild},
  V. 2006, \mnras, 371, 495

\bibitem[{{Bouch{\'e}} {et~al.}(2007){Bouch{\'e}}, {Murphy}, {P{\'e}roux},
  {Davies}, {Eisenhauer}, {F{\"o}rster Schreiber}, \& {Tacconi}}]{bouche2007ha}
{Bouch{\'e}}, N., {Murphy}, M.~T., {P{\'e}roux}, C., {Davies}, R.,
  {Eisenhauer}, F., {F{\"o}rster Schreiber}, N.~M., \& {Tacconi}, L. 2007,
  \apjl, 669, L5

\bibitem[{{Butcher} \& {Oemler}(1984)}]{butcheroemler}
{Butcher}, H., \& {Oemler}, Jr., A. 1984, \apj, 285, 426

\bibitem[{{Charlton} {et~al.}(2003){Charlton}, {Ding}, {Zonak}, {Churchill},
  {Bond}, \& {Rigby}}]{charltonetal03}
{Charlton}, J.~C., {Ding}, J., {Zonak}, S.~G., {Churchill}, C.~W., {Bond},
  N.~A., \& {Rigby}, J.~R. 2003, \apj, 589, 111

\bibitem[{{Chen} {et~al.}(2010){Chen}, {Helsby}, {Gauthier}, {Shectman},
  {Thompson}, \& {Tinker}}]{chen2010lowz}
{Chen}, H., {Helsby}, J.~E., {Gauthier}, J., {Shectman}, S.~A., {Thompson},
  I.~B., \& {Tinker}, J.~L. 2010, ArXiv e-prints

\bibitem[{{Chen} \& {Tinker}(2008)}]{chenandtinker2008}
{Chen}, H., \& {Tinker}, J.~L. 2008, \apj, 687, 745

\bibitem[{{Churchill} {et~al.}(2000){Churchill}, {Mellon}, {Charlton},
  {Jannuzi}, {Kirhakos}, {Steidel}, \& {Schneider}}]{churchill_zoo}
{Churchill}, C.~W., {Mellon}, R.~R., {Charlton}, J.~C., {Jannuzi}, B.~T.,
  {Kirhakos}, S., {Steidel}, C.~C., \& {Schneider}, D.~P. 2000, \apj, 543, 577

\bibitem[{{Churchill} {et~al.}(1999){Churchill}, {Rigby}, {Charlton}, \&
  {Vogt}}]{churchill_weakmgii}
{Churchill}, C.~W., {Rigby}, J.~R., {Charlton}, J.~C., \& {Vogt}, S.~S. 1999,
  \apjs, 120, 51

\bibitem[{{Coil} {et~al.}(2004){Coil}, {Newman}, {Kaiser}, {Davis}, {Ma},
  {Kocevski}, \& {Koo}}]{coil_deep2}
{Coil}, A.~L., {Newman}, J.~A., {Kaiser}, N., {Davis}, M., {Ma}, C.,
  {Kocevski}, D.~D., \& {Koo}, D.~C. 2004, \apj, 617, 765

\bibitem[{{Davis} {et~al.}(2003){Davis}, {Faber}, {Newman}, {Phillips},
  {Ellis}, {Steidel}, {Conselice}, {Coil}, {Finkbeiner}, {Koo}, {Guhathakurta},
  {Weiner}, {Schiavon}, {Willmer}, {Kaiser}, {Luppino}, {Wirth}, {Connolly},
  {Eisenhardt}, {Cooper}, \& {Gerke}}]{deep2_1}
{Davis}, M., {et~al.} 2003, in Presented at the Society of Photo-Optical
  Instrumentation Engineers (SPIE) Conference, Vol. 4834, Society of
  Photo-Optical Instrumentation Engineers (SPIE) Conference Series, ed.
  {P.~Guhathakurta}, 161--172

\bibitem[{{Davis} {et~al.}(2007){Davis}, {Guhathakurta}, {Konidaris}, {Newman},
  {Ashby}, {Biggs}, {Barmby}, {Bundy}, {Chapman}, {Coil}, {Conselice},
  {Cooper}, {Croton}, {Eisenhardt}, {Ellis}, {Faber}, {Fang}, {Fazio},
  {Georgakakis}, {Gerke}, {Goss}, {Gwyn}, {Harker}, {Hopkins}, {Huang},
  {Ivison}, {Kassin}, {Kirby}, {Koekemoer}, {Koo}, {Laird}, {Le Floc'h}, {Lin},
  {Lotz}, {Marshall}, {Martin}, {Metevier}, {Moustakas}, {Nandra}, {Noeske},
  {Papovich}, {Phillips}, {Rich}, {Rieke}, {Rigopoulou}, {Salim},
  {Schiminovich}, {Simard}, {Smail}, {Small}, {Weiner}, {Willmer}, {Willner},
  {Wilson}, {Wright}, \& {Yan}}]{deep2_2}
{Davis}, M., {et~al.} 2007, \apjl, 660, L1

\bibitem[{{Dekel} {et~al.}(2009){Dekel}, {Birnboim}, {Engel}, {Freundlich},
  {Goerdt}, {Mumcuoglu}, {Neistein}, {Pichon}, {Teyssier}, \& {Zinger}}]{dekel}
{Dekel}, A., {et~al.} 2009, \nat, 457, 451

\bibitem[{{Faucher-Giguere} \& {Keres}(2010)}]{faucher_coldflows}
{Faucher-Giguere}, C., \& {Keres}, D. 2010, ArXiv e-prints

\bibitem[{{Faucher-Gigu{\`e}re} {et~al.}(2009){Faucher-Gigu{\`e}re}, {Lidz},
  {Zaldarriaga}, \& {Hernquist}}]{giguereuvbg}
{Faucher-Gigu{\`e}re}, C., {Lidz}, A., {Zaldarriaga}, M., \& {Hernquist}, L.
  2009, \apj, 703, 1416

\bibitem[{{Gauthier} {et~al.}(2009){Gauthier}, {Chen}, \& {Tinker}}]{gauthier}
{Gauthier}, J., {Chen}, H., \& {Tinker}, J.~L. 2009, \apj, 702, 50

\bibitem[{{Gerke} {et~al.}(2007{\natexlab{a}}){Gerke}, {Newman}, {Faber},
  {Cooper}, {Croton}, {Davis}, {Willmer}, {Yan}, {Coil}, {Guhathakurta}, {Koo},
  \& {Weiner}}]{gerke}
{Gerke}, B.~F., {et~al.} 2007{\natexlab{a}}, \mnras, 376, 1425

\bibitem[{{Gerke} {et~al.}(2007{\natexlab{b}}){Gerke}, {Newman}, {Faber},
  {Cooper}, {Croton}, {Davis}, {Willmer}, {Yan}, {Coil}, {Guhathakurta}, {Koo},
  \& {Weiner}}]{gerke2007}
---. 2007{\natexlab{b}}, \mnras, 376, 1425

\bibitem[{Jakobsson {et~al.}(2004)Jakobsson, Hjorth, Fynbo, Weidinger,
  Gorosabel, Ledoux, Watson, Bjornsson, Gudmundsson, Wijers, M{\o}ller,
  Pedersen, Sollerman, Henden, Jensen, Gilmore, Kilmartin, Levan, Cer\'on,
  Castro-Tirado, Fruchter, Kouveliotou, Masetti, \& Tanvir}]{jakobssongrbpair}
Jakobsson, P., {et~al.} 2004, A\&A, 427, 785

\bibitem[{{Jarosik} {et~al.}(2010){Jarosik}, {Bennett}, {Dunkley}, {Gold},
  {Greason}, {Halpern}, {Hill}, {Hinshaw}, {Kogut}, {Komatsu}, {Larson},
  {Limon}, {Meyer}, {Nolta}, {Odegard}, {Page}, {Smith}, {Spergel}, {Tucker},
  {Weiland}, {Wollack}, \& {Wright}}]{wmap7}
{Jarosik}, N., {et~al.} 2010, ArXiv e-prints

\bibitem[{{Kacprzak} {et~al.}(2010){Kacprzak}, {Churchill}, {Ceverino},
  {Steidel}, {Klypin}, \& {Murphy}}]{kacprzak2010sim}
{Kacprzak}, G.~G., {Churchill}, C.~W., {Ceverino}, D., {Steidel}, C.~C.,
  {Klypin}, A., \& {Murphy}, M.~T. 2010, \apj, 711, 533

\bibitem[{{Kere{\v s}} {et~al.}(2005){Kere{\v s}}, {Katz}, {Weinberg}, \&
  {Dav{\'e}}}]{keres}
{Kere{\v s}}, D., {Katz}, N., {Weinberg}, D.~H., \& {Dav{\'e}}, R. 2005,
  \mnras, 363, 2

\bibitem[{{Lemaux} {et~al.}(2009){Lemaux}, {Lubin}, {Sawicki}, {Martin},
  {Lagattuta}, {Gal}, {Kocevski}, {Fassnacht}, \& {Squires}}]{lemaux}
{Lemaux}, B.~C., {et~al.} 2009, \apj, 700, 20

\bibitem[{{Lundgren} {et~al.}(2011){Lundgren}, {Wake}, {Padmanabhan}, {Coil},
  \& {York}}]{lundgren}
{Lundgren}, B., {Wake}, D., {Padmanabhan}, N., {Coil}, A., \& {York}, D. 2011,
  ArXiv e-prints

\bibitem[{{Marshall} {et~al.}(2008){Marshall}, {Burles}, {Thompson},
  {Shectman}, {Bigelow}, {Burley}, {Birk}, {Estrada}, {Jones}, {Smith},
  {Kowal}, {Castillo}, {Storts}, \& {Ortiz}}]{Marshall}
{Marshall}, J.~L., {et~al.} 2008, in Society of Photo-Optical Instrumentation
  Engineers (SPIE) Conference Series, Vol. 7014, Society of Photo-Optical
  Instrumentation Engineers (SPIE) Conference Series

\bibitem[{{M{\'e}nard} {et~al.}(2009){M{\'e}nard}, {Wild}, {Nestor}, {Quider},
  \& {Zibetti}}]{menardo2}
{M{\'e}nard}, B., {Wild}, V., {Nestor}, D., {Quider}, A., \& {Zibetti}, S.
  2009, ArXiv e-prints

\bibitem[{{Moustakas} {et~al.}(2006){Moustakas}, {Kennicutt}, \&
  {Tremonti}}]{moustakassfr2006}
{Moustakas}, J., {Kennicutt}, Jr., R.~C., \& {Tremonti}, C.~A. 2006, \apj, 642,
  775

\bibitem[{{Nestor} {et~al.}(2010){Nestor}, {Johnson}, {Wild}, {M{\'e}nard},
  {Turnshek}, {Rao}, \& {Pettini}}]{nestor2010strong}
{Nestor}, D.~B., {Johnson}, B.~D., {Wild}, V., {M{\'e}nard}, B., {Turnshek},
  D.~A., {Rao}, S., \& {Pettini}, M. 2010, ArXiv e-prints

\bibitem[{{Nestor} {et~al.}(2005){Nestor}, {Turnshek}, \& {Rao}}]{nestor2005}
{Nestor}, D.~B., {Turnshek}, D.~A., \& {Rao}, S.~M. 2005, \apj, 628, 637

\bibitem[{{Norberg} {et~al.}(2002){Norberg}, {Baugh}, {Hawkins}, {Maddox},
  {Madgwick}, {Lahav}, {Cole}, {Frenk}, {Baldry}, {Bland-Hawthorn}, {Bridges},
  {Cannon}, {Colless}, {Collins}, {Couch}, {Dalton}, {De Propris}, {Driver},
  {Efstathiou}, {Ellis}, {Glazebrook}, {Jackson}, {Lewis}, {Lumsden},
  {Peacock}, {Peterson}, {Sutherland}, \& {Taylor}}]{norberg_2df}
{Norberg}, P., {et~al.} 2002, \mnras, 332, 827

\bibitem[{Prochter {et~al.}(2006)Prochter, Prochaska, , \&
  Burles}]{prochtersdss2006}
Prochter, G.~E., Prochaska, J.~X., , \& Burles, S.~M. 2006, The Astrophysical
  Journal, 639, 766

\bibitem[{{Prochter} {et~al.}(2006){Prochter}, {Prochaska}, {Chen}, {Bloom},
  {Dessauges-Zavadsky}, {Foley}, {Lopez}, {Pettini}, {Dupree}, \&
  {Guhathakurta}}]{prochtergrbsightlines}
{Prochter}, G.~E., {et~al.} 2006, \apjl, 648, L93

\bibitem[{{Rauch} {et~al.}(2002){Rauch}, {Sargent}, {Barlow}, \&
  {Simcoe}}]{rauch}
{Rauch}, M., {Sargent}, W.~L.~W., {Barlow}, T.~A., \& {Simcoe}, R.~A. 2002,
  \apj, 576, 45

\bibitem[{{Steidel}(1995)}]{steidel_scaling}
{Steidel}, C.~C. 1995, in QSO Absorption Lines, ed. {G.~Meylan}, 139--+

\bibitem[{{Steidel} {et~al.}(1994){Steidel}, {Dickinson}, \&
  {Persson}}]{steidel}
{Steidel}, C.~C., {Dickinson}, M., \& {Persson}, S.~E. 1994, \apjl, 437, L75

\bibitem[{{Steidel} {et~al.}(2002){Steidel}, {Kollmeier}, {Shapley},
  {Churchill}, {Dickinson}, \& {Pettini}}]{mgii_rotation}
{Steidel}, C.~C., {Kollmeier}, J.~A., {Shapley}, A.~E., {Churchill}, C.~W.,
  {Dickinson}, M., \& {Pettini}, M. 2002, \apj, 570, 526

\bibitem[{{Steidel} \& {Sargent}(1992)}]{steidelandsargent}
{Steidel}, C.~C., \& {Sargent}, W.~L.~W. 1992, \apjs, 80, 1

\bibitem[{{Tinker} \& {Chen}(2010)}]{chenandtinker2009}
{Tinker}, J.~L., \& {Chen}, H. 2010, \apj, 709, 1

\bibitem[{{Weiner} {et~al.}(2009){Weiner}, {Coil}, {Prochaska}, {Newman},
  {Cooper}, {Bundy}, {Conselice}, {Dutton}, {Faber}, {Koo}, {Lotz}, {Rieke}, \&
  {Rubin}}]{ubiquitous2009}
{Weiner}, B.~J., {et~al.} 2009, \apj, 692, 187

\bibitem[{{Yan} {et~al.}(2009){Yan}, {Newman}, {Faber}, {Coil}, {Cooper},
  {Davis}, {Weiner}, {Gerke}, \& {Koo}}]{yan_poststarburst}
{Yan}, R., {et~al.} 2009, \mnras, 398, 735

\bibitem[{{Zhu} {et~al.}(2009){Zhu}, {Moustakas}, \& {Blanton}}]{zhuoii2009}
{Zhu}, G., {Moustakas}, J., \& {Blanton}, M.~R. 2009, \apj, 701, 86

\bibitem[{{Zibetti} {et~al.}(2007){Zibetti}, {M{\'e}nard}, {Nestor}, {Quider},
  {Rao}, \& {Turnshek}}]{zibetti2007stacking}
{Zibetti}, S., {M{\'e}nard}, B., {Nestor}, D.~B., {Quider}, A.~M., {Rao},
  S.~M., \& {Turnshek}, D.~A. 2007, \apj, 658, 161

\end{thebibliography}
\bibliographystyle{apj}
\end{document}